\documentclass[11pt]{article}
\pdfoutput=1
\usepackage{authblk}
\usepackage{palatino}
\usepackage[fleqn]{amsmath}
\usepackage{amsmath}
\usepackage{amsthm}
\usepackage{amssymb}
\usepackage{boxedminipage}
\usepackage[mathcal]{euscript}
\usepackage[all]{xy}
\usepackage{graphics}
\usepackage{graphicx}
\usepackage{subfig}
\usepackage{multirow}
\usepackage{url}

\newtheorem{theorem}{\bf Theorem}[section]

\newtheorem{lemma}[theorem]{\bf Lemma}

\newcommand{\optcost} {{\sf cost^*}}
\newcommand{\opt} {{\sf Opt}}

\newcommand{\ttmcP} {\tenT[P]}
\newcommand{\lbl} {{\sf lbl}}
\newcommand{\rt} {{\sf root}}
\newcommand{\In}[1] {{\mathcal In}(#1)}
\newcommand{\Out}[1] {{\mathcal Out}(#1)}
\newcommand{\argmin} {{\rm argmin}}

\newcommand{\dvol} {{\sf dvol}}

\newcommand{\vect}[1] {\overrightarrow{#1}}

\newcommand{\tenT} {{\mathcal T}}
\newcommand{\tenX} {{\mathcal X}}

\newcommand{\tenZ} {{\mathcal Z}}

\newcommand{\gcore} {{\mathcal G}}

\newcommand{\matF} {\mathbf{F}}
\newcommand{\tmatF} {\mathbf{F}^T}
\newcommand{\matA} {\mathbf{A}}

\newcommand{\newF} {\widetilde{\matF}}
\newcommand{\tnewF} {\widetilde{\matF}^T}
\newcommand{\newg} {\tilde{\gcore}}

\newcommand{\myfloor}[1] {\lfloor #1 \rfloor}

\newcommand{\ucar} {U{\c{c}}ar}
\newcommand{\mnuf}[1] {#1_{(n)}}

\newcommand{\eat}[1] {}

\newcommand{\pg} {{\sf pg}}
\newcommand{\rg} {{\sf rg}^*}
\newcommand{\vol} {{\sf vol}}
\newcommand{\gpar} {g_{\sf par}}
\newcommand{\subT}[1] {H(#1)}
\renewcommand{\paragraph}[1] {\vspace*{0.3cm}\noindent{\it #1}: }

\begin{document}
\title{On Optimizing Distributed Tucker Decomposition for Dense Tensors}
\author{Venkatesan T. Chakaravarthy} 
\author{Jee W. Choi} 
\author{Douglas J. Joseph} 
\author{Xing Liu}
\author{Prakash Murali} 
\author{Yogish Sabharwal} 
\author{Dheeraj Sreedhar}

\affil{
	IBM Research\\
	\it{\{vechakra,prakmura,ysabharwal,dhsreedh\}@in.ibm.com}\\
	\it{\{jwchoi,djoseph,xliu\}@us.ibm.com}
}
\date{~}

\maketitle     

\begin{abstract}
The Tucker decomposition expresses a given tensor as the product of a small core tensor and a set of factor matrices.
Apart from providing data compression, the construction is useful in performing analysis 
such as principal component analysis (PCA) and finds applications in diverse domains such as signal processing, computer vision and text analytics. 
Our objective is to develop an efficient distributed implementation for the case of dense tensors.
The implementation is based on the HOOI (Higher Order Orthogonal Iterator) procedure, 
wherein the tensor-times-matrix product forms the core routine.
Prior work have proposed heuristics for reducing the computational load and communication volume incurred by the routine.
We study the two metrics in a formal and systematic manner, 
and design strategies that are optimal under the two fundamental metrics.
Our experimental evaluation on a large benchmark of tensors
shows that the optimal strategies provide significant reduction in load and volume
compared to prior heuristics, and provide up to $7$x speed-up in the overall running time.
\end{abstract}

\section{Introduction}
\begin{figure}
\centerline{
\includegraphics[width=4.0in]{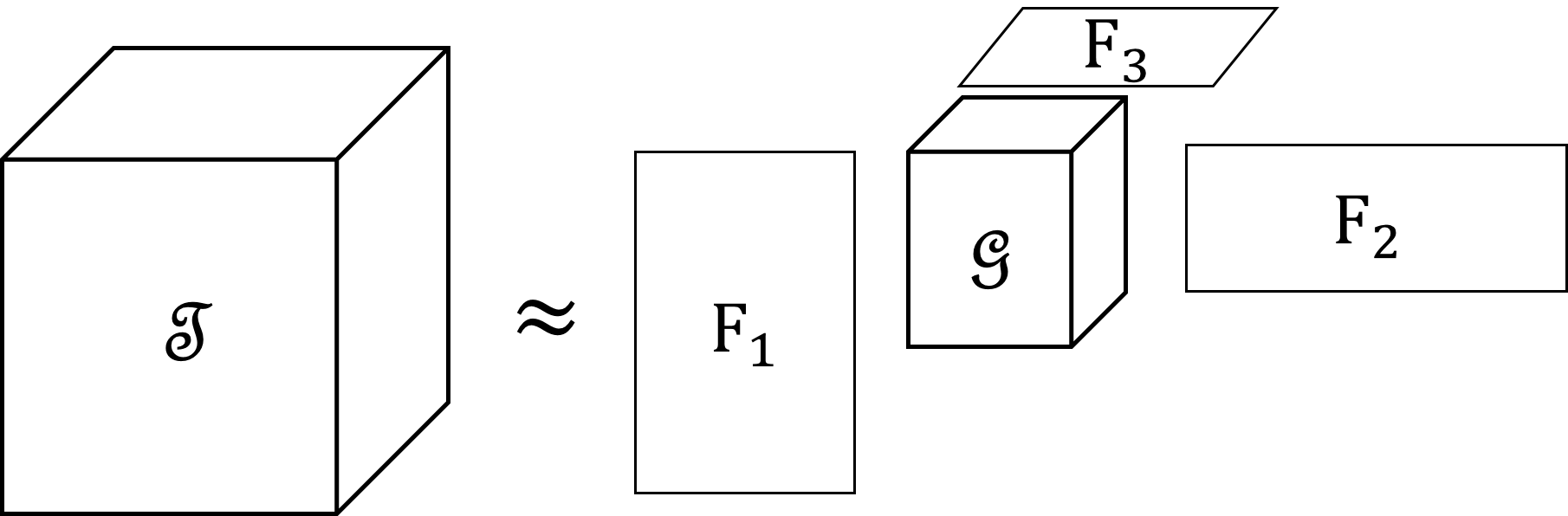}
}
\caption{Illustration for Tucker decomposition on a $3$-dimensional tensor: $\tenT$ - input tensor, $\gcore$ - core tensor,
$\matF_n$  - factor matrices.}
	\label{fig:tucker}
\end{figure}

Tensors are the higher dimensional analogues of matrices. While matrices represent two-dimensional data,
tensors are useful in representing data in three or higher dimensions.
Tensors have been studied extensively via generalizing concepts pertaining to matrices.
The Tucker decomposition \cite{tucker} is a prominent construction
that extends the singular value decomposition (SVD) to the setting of tensors. 
Given an $N$-dimensional tensor $\tenT$, the decomposition approximately expresses 
the tensor as the product of a small $N$-dimensional core tensor $\gcore$ and a set of $N$ factor matrices, 
one along each dimension (or {\em mode}); see Figure \ref{fig:tucker} for an illustration.
The core is much smaller than the original tensor leading to data compression.
Prior work \cite{kolda-ipdps} has reported compression rates to the tune of $5000$ on large real tensors.
Apart from data compression, the decomposition is also useful in analysis such as PCA,
and finds applications in diverse domains such as computer vision \cite{s229} and  signal processing \cite{s173}.
A detailed discussion on the topic can be found in the excellent survey, by Kolda and Bader \cite{survey}.

Tucker decomposition has been well-studied in sequential, 
shared memory and distributed settings for both dense and sparse tensors (e.g., \cite{kolda-ipdps,ucar-icpp,kolda-icdm}).
Our objective is to develop an optimized implementation for dense tensors on distributed memory systems.
The implementation is based on the popular STHOSVD/HOOI procedures. 
The STHOSVD (Sequentially Truncated Higher Order SVD) \cite{vanv} is used to produce an initial decomposition.
The HOOI (Higher Order Orthogonal Iterator) \cite{kolda7} procedure transforms any given decomposition
to a new decomposition with the same core size, but with reduced error.
The procedure is applied iteratively so as to reduce the error monotonically across the iterations.
We focus on the latter HOOI procedure which is invoked multiple times. 
The tensor-times-matrix product (TTM) component forms the core module of the procedure. 
Prior work has proposed heuristic schemes for reducing the computational load and communication volume of the component.
Our objective is to enhance the performance by constructing schemes which are {\em optimal} in these two fundamental metrics. 

\subsection*{Prior Work}
\paragraph{Heuristics for computational load} 
The TTM component comprises of a set of tensor-times-matrix multiplication operations.
Based on the observation that the operations can be rearranged and reused in multiple ways,
prior work has proposed heuristics for reducing the computational load.
A naive scheme for implementing the component performs $N(N-1)$ TTM operations. 
Baskaran et al. \cite{baskaran} focused on reducing the number of TTM operations,
and proposed a scheme with (approximately) $N^2/2$ operations,
which was further improved to $N\log N$ by Kaya and {\ucar} \cite{ucar-report}.
However, minimizing the number of TTM operations is not sufficient
and it is crucial to consider the cost of the operations, especially in the context of dense tensors.
Austin et al. \cite{kolda-ipdps} measure the cost in terms of the number of floating point operations (FLOP).
They empirically showed that the performance of the navie scheme
can be improved by permuting (ordering) the modes of the input tensor 
and proposed a greedy heuristic for mode ordering.
A similar heuristic is given by Vannieuwenhoven et al. \cite{vanv}.

\paragraph{Heuristics for communication volume} 
Austin et al. \cite{kolda-ipdps} presented the first distributed implementation of HOOI.
They distribute the tensors among the processors using a Cartesian parallel distribution 
which generalizes the block distribution technique used in the context of matrices. 
The processors are arranged in the form of an $N$-dimensional grid and 
a tensor is partitioned into blocks by imposing the grid on the tensor;
the blocks are then assigned to the processors.
They showed that the communication volume is determined by the choice of the grid,
and presented an empirical evaluation of the effect of the grid on the communication volume.

\subsection*{Our Contributions}
Our objective is to enhance the performance of distributed Tucker decomposition for dense tensors by designing 
optimal schemes for the two metrics and we make the following contributions.
\begin{itemize}
\item
{\it Optimal TTM-trees:}
As observed in prior work \cite{ucar-report}, the different TTM schemes can be conveniently represented in the form of trees, called TTM-trees.
We present an efficient algorithm for constructing the optimal TTM-tree,
the one having the least computational load, measured in terms of number of floating operations (FLOP). 

\item 
{\it Dynamic Gridding:} 
Prior work uses a static gridding scheme, wherein the same grid is used for distributing 
the tensors arising in the different TTM operations. 
We propose the concept of {\em dynamic gridding} that uses different grids tailored for the different operations,
leading to significant reduction in communication volume, even when compared to the optimal static grids.

\item
{\it Optimal Dynamic Gridding:}
We present an efficient algorithm for finding the dynamic gridding scheme achieving the optimal communication volume.

%\item
%{\it Experimental Evaluation:}
%We present an experimental evaluation involving a large number of tensors
%having different dimension sizes that mimic typical real-life tensors.
%The study shows that the combination of optimal trees and the dynamic gridding scheme offer
%significant reduction in computational load and communication volume,
%resulting in up to $7$x factor improvement in overall running time,
%compared to prior heuristics.
\end{itemize}
Our distributed implementation builds on the work of Austin et al. \cite{kolda-ipdps}
and incorporates the optimal schemes described above.
%At its core, the TTM component involves dense matrix-matrix multiplications.
%Consequently, the running time is independent of the actual values of the tensor elements.
%Instead, the execution time of any implementation of the component depends crucially
%on the metadata - the dimension lengths of the input and the core tensors.
%We exploit this observation to 
We setup a large benchmark consisting of about $1700$ tensors whose metadata are derived from real-life tensors.
Furthermore, we also include a set of tensors with metadata derived from simulations in combustion science.
Our experimental evaluation on the above benchmark demonstrates that the combination of optimal trees and the dynamic gridding scheme offers 
significant reduction in computational load and communication volume, 
resulting in up to $7$-factor improvement in overall execution time, compared to prior heuristics.
To the best of our knowledge, our study is the first to consider optimal algorithms for the Tucker decomposition.

We note that prior work \cite{vanv,kolda-ipdps} has provided evidence that STHOSVD may be sufficient for particular application domains. 
They present experimental evaluations on a sample of tensors arising in image processing and combustion science showing that for these tensors, 
STHOSVD is sufficient and HOOI does not provide significant error reduction. 
Our optimizations on HOOI would be useful for other tensors/domains where HOOI provides error reduction over STHOSVD. 
Furthermore, the ideas developed in this paper can be recast and used for improving STHOSVD as well.

\subsection*{Related Work}
Tucker decomposition has been studied in sequential and parallel 
settings for dense and sparse tensors. 
For dense tensors, the MATLAB Tensor 
Toolbox provides a sequential implementation \cite{TTB_Software2}. 
Zhou et al. \cite{zhou} proposed a randomized algorithm 
for the case where the tensor fits in the physical memory of a single machine. 
Li et al. \cite{vuduc} proposed performance enhancements for a single TTM operation
and their techniques can be incorporated within our framework.
Austin et al. \cite{kolda-ipdps} 
described the first implementation for distributed memory systems, 
wherein they proposed heuristics for mode ordering
and experimentally demonstrated the effect of grid selection on communication time. 
For sparse tensors, sequential \cite{kolda-icdm}, shared memory \cite{baskaran} and 
distributed implementations \cite{ucar-icpp} are known. 
Other tensor decompositions have also been considered (see \cite{survey}).
In particular, CP decomposition which generalizes the concept of rank 
factorization has been well studied (e.g. \cite{cp3}).
We refer to survey by Kolda and Bader \cite{survey} for a detailed treatment of tensor decompositions.

\section{Tucker Decomposition}
In this section, we briefly discuss tensor concepts pertinent to our problem,
and then, describe the Tucker decomposition and the HOOI procedure.

\subsection{Preliminaries}
\paragraph{Fibers}
Consider an $N$-dimensional tensor $\tenT$ of size $L_1\times L_2\times \cdots \times L_N$.
Let $|\tenT|$ denote the cardinality (number of elements) of the tensor.
The elements of $\tenT$ can be canonically indexed by a coordinate vector of the form $\langle l_1, l_2, \ldots, l_N\rangle$,
where each index $l_n$ belongs to $[1, L_n]$, for all modes $1\leq n\leq N$.
A {\em mode-$n$ fiber} $\vect{x}$ is a vector of length $L_n$, containing all the elements
that differ on the $n$th coordinate, but agree on all the other coordinates,
i.e., $\langle l_1, \ldots, l_{n-1}, *, l_{n+1}, \ldots l_N\rangle$.
The number of mode-$n$ fibers is  $|\tenT|/L_n$.
In the analogous case of matrices, two types of fibers can be found: row vectors and column vectors.

\paragraph{Tensor Unfolding}
The tensor $\tenT$ is stored as a matrix and there are $N$ different matrix layouts are possible, called the unfoldings.
The {\em mode-$n$ unfolding} of $\tenT$ refers to the matrix whose columns are the mode-$n$ fibers of the tensor.
The columns are arranged in a lexicographic order (the details are not crucial for our discussion).
This matrix is of size $L_n \times (|\tenT|/L_n)$, and we denote it as $\mnuf{T}$.

%Figure~\ref{fig:mat-ex} provides an illustration of the concept. 
%\begin{figure}
%\begin{small}
%\begin{tabular}{c}
%$\begin{bmatrix}
%	\ele{1,1,1} & \ele{1,2,1} &  \ele{1,1,2}  & \ele{1,2,2}\\
%	\ele{2,1,1} & \ele{2,2,1} &  \ele{2,1,2}  & \ele{2,2,2}\\
%	\ele{3,1,1} & \ele{3,2,1} &  \ele{3,1,2}  & \ele{3,2,2}
%\end{bmatrix}$
%\\
%\\
%$\begin{bmatrix}
%	\ele{1,1,1} & \ele{2,1,1} &  \ele{3,1,1}  & \ele{1,1,1}  & \ele{2,1,2}  & \ele{3,1,2}\\
%	\ele{1,2,1} & \ele{2,2,1} &  \ele{3,2,1}  & \ele{1,2,1}  & \ele{2,2,2}  & \ele{3,2,2}
%\end{bmatrix}$ \\
%\\
%$\begin{bmatrix}
%	\ele{1,1,1} & \ele{2,1,1} &  \ele{3,1,1}  & \ele{1,2,1}  & \ele{2,2,1}  & \ele{3,2,1}\\
%	\ele{1,1,2} & \ele{2,1,2} &  \ele{3,1,2}  & \ele{1,2,2}  & \ele{2,2,2}  & \ele{3,2,2}
%\end{bmatrix}$
%\end{tabular}
%\end{small}
%\caption{Unfoldings of a $3\times 2\times 2$ tensor along modes 1, 2 and 3. Instead of actual values, the elements are depicted via their coordinate vectors.}
%\label{fig:mat-ex}
%\end{figure}

\paragraph{Tensor-Times-Matrix Multiplication (TTM)}
For any mode $n$, the tensor $\tenT$ can be multiplied by a matrix $\matA$ along mode $n$, provided $\matA$ has size $K\times L_n$, for some $K$;
the operation is denoted $\tenZ =\tenT\times_n \matA$.
Conceptually, the operation applies the linear transformation $\matA$ to all the mode-$n$ fibers. 
It is realized via the matrix-matrix multiplication $\matA \times \mnuf{T}$, 
and taking the output matrix to be the mode-$n$ unfolding of $\tenZ$.
While the length along mode $n$ changes from $L_n$ to $K$, 
the number of fibers and the lengths along other modes remains the same.
Thus, $\tenZ$ has cardinality $K\cdot (|\tenT|/L_n)$ and size $L_1\times \cdots \times L_{n-1}\times K\times L_{n+1}\times L_N$.

%The {\em Tensor-Times-Matrix} (TTM) operation refers to multiplying the tensor $\tenT$ by a matrix $\matA$ along any mode $n$ 
%and it is denoted $\tenZ = \tenT\times_n \matA$; for this, $\matA$ must have size $K\times L_n$, for some $K$.
%Conceptually, the operation applies the linear transformation $\matA$ to all the mode-$n$ fibers changing their length from $L_n$ to $K$.
%Namely, we derive $\tenZ$ by replacing each mode-$n$ fiber $\vect{x}$ by the transformed fiber $\vect{y}$, 
%given by the matrix-vector product $\vect{y}=\matA\cdot \vect{x}$. 
%While the length along mode $n$ changes from $L_n$ to $K$,
%the number of fibers and the lengths along other modes remains the same,
%i.e., $\tenZ$ has size $L_1\times \cdots \times L_{n-1}\times K\times L_{n+1}\times L_N$.
%%and cardinality $|\tenZ|=(K/L_n)|\tenT|$.
%In terms of implementation, the operation is realized via the matrix-matrix multiplication
%$\mnuf{Z}=\matA \times \mnuf{T}$, where $\mnuf{T}$ and $\mnuf{Z}$ are the mode-$n$ unfoldings of $\tenT$ and $\tenZ$, respectively.

\paragraph{TTM-Chain}
The {\em TTM-chain} operation refers to multiplying $\tenT$ along multiple distinct modes.
For two modes $n_1$ and $n_2$ and matrices $\matA_1$ and $\matA_2$,
we first multiply $\tenT$ by $\matA_1$ along mode $n_1$, and then multiply the output tensor along mode $n_2$ by $\matA_2$.
An important property of the operation is commutativity \cite{kolda7}, namely the two TTM operations can be performed in any order:
	$(\tenT \times_{n_1} \matA_1)\times_{n_2} \matA_2 = (\tenT\times_{n_2} \matA_2)\times_{n_1} \matA_1$.
In general, for a subset of distinct modes $S=\{n_1, n_2, \ldots, n_r\}$,
and matrices $\matA_1, \matA_2, \ldots, \matA_r$, where $\matA_j$ has size $K_j \times L_{n_j}$,
the output is a tensor $\tenZ = \tenT \times_{n_1}\matA_1 \times \cdots\times_{n_r} \matA_{n_r}$.
The length of $\tenZ$ remains the same as $\tenT$ along  modes not belonging to $S$,
and changes to $K_j$, for all $n_j\in S$.
Commutativity implies that the multiplications can be performed in any order.

\subsection{Tucker Decomposition and HOOI}
The Tucker decomposition of $\tenT$ approximates the tensor
as the product of a {\em core tensor} $\gcore$ of size $K_1\times K_2\times \cdots \times K_N$, with each $K_n\leq L_n$,
and a set of {\em factor matrices} $\matF_1, \matF_2, \ldots, \matF_N$:
$\tenT \approx \tenZ = \gcore\times_1 \matF_1\times_2 \matF_2 \times \cdots \times_N \matF_N$.
The factor matrix $\matF_n$ has size $L_n \times K_n$.
The decomposition compresses the length of $\tenT$ along each mode from $L_n$ to $K_n$.
We write the decomposition as $\{\gcore;\matF_1,\matF_2,\ldots,\matF_N\}$.
The error of the decomposition is measured by comparing the recovered tensor $\tenZ$ and the input tensor $\tenT$
under the normalized root mean square metric.

The HOOI procedure \cite{kolda7} transforms a given decomposition into a new decomposition having the same core size, but with reduced the error.
Given an initial decomposition, the procedure can be invoked repeatedly
to reduce the error monotonically, until a desired convergence is achieved.
An initial decomposition can be found using methods such as STHOSVD \cite{vanv}.

\begin{figure}
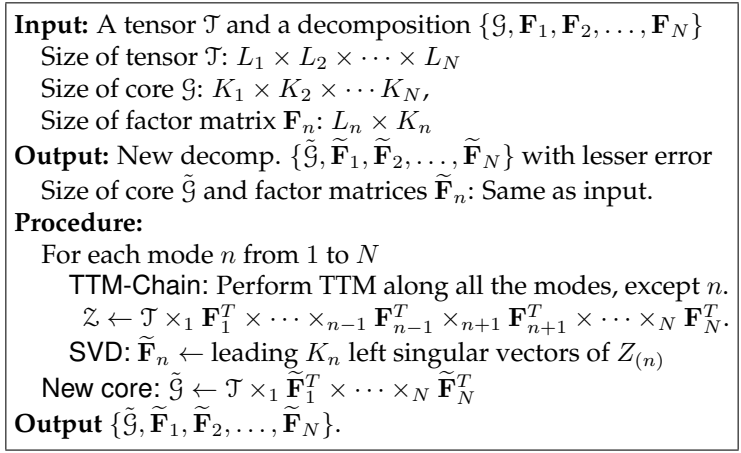

\begin{center}
\begin{boxedminipage}{\hsize}
\begin{small}
\begin{tabbing}
xx\=xx\=xx\=xx\=xx\=xx\=xx\=\kill
\textbf{Input:} A tensor $\tenT$ and a decomposition $\{\gcore, \matF_1, \matF_2, \ldots, \matF_N\}$ \\
\> Size of tensor $\tenT$: $L_1\times L_2\times \cdots\times L_N$\\
\> Size of core $\gcore$: $K_1\times K_2\times \cdots K_N$,\\
\> Size of factor matrix $\matF_n$: $L_n\times K_n$\\
\textbf{Output:} New decomp.  $\{\newg, \newF_1, \newF_2, \ldots, \newF_N\}$ with lesser error\\
\> Size of core $\newg$ and factor matrices $\newF_n$: Same as input.\\
\textbf{Procedure:}\\
\> For each mode $n$ from $1$ to $N$\\
\> \> {\sf TTM-Chain:} Perform TTM along all the modes, except $n$.\\
\> \> ~~$\tenZ \leftarrow \tenT \times_1 \tmatF_1\times \cdots \times_{n-1} \tmatF_{n-1}\times_{n+1} \tmatF_{n+1}\times \cdots \times_N \tmatF_N$.\\
\> \> {\sf SVD:} $\newF_n \leftarrow$ leading $K_n$ left singular vectors of $\mnuf{Z}$\\
\> {\sf New core:} $\newg \leftarrow \tenT\times_1 \tnewF_1\times \cdots \times_N \tnewF_N$\\
\textbf{Output} $\{\newg, \newF_1, \newF_2, \ldots, \newF_N\}$.
\end{tabbing}
\end{small}
\end{boxedminipage}
\end{center}
\caption{HOOI Procedure}
\label{fig:hooi}
\end{figure}

The HOOI procedure (a single invocation), shown in Figure \ref{fig:hooi}, 
takes as input the tensor $\tenT$, and a decomposition $\{\gcore, \matF_1, \matF_2, \ldots, \matF_N\}$ with core size $K_1\times K_2\times \cdots \times K_N$.
It produces a new decomposition $\{\newg, \newF_1, \newF_2, \ldots, \newF_N\}$ with lesser error, but having the same core and factor matrix sizes.
For computing each new factor matrix $\newF_n$, the procedure utilizes the alternating least squares paradigm and works in two steps.
First, it performs a TTM-chain operation by skipping mode $n$ and multiplying $\tenT$ by the transposes of all the other factor matrices $\matF_j$ (with $j\neq n$) 
and obtains a tensor $\tenZ$.  The tensor $\tenZ$ has length compressed from $L_j$ to $K_j$ along all modes $j\neq n$.
In the next step, it performs an SVD on $\mnuf{Z}$, the mode-$n$ unfolding of the $\tenZ$.
The new factor matrix $\newF_n$ is obtained by arranging the leading $K_n$ singular vectors as columns.
Once all the new factor matrices are computed, the new core tensor is computed.

Figure \ref{fig:load-ex} (a) depicts the process in the form of a tree.
The root represents the input tensor $\tenT$, each node with label $n$ represents multiplication along mode $n$,
and each leaf represents a new factor matrix.
For dense tensors, the SVD operations tend to be inexpensive (see \cite{kolda-ipdps}).
Therefore, we focus on optimizing the TTM component comprising of the $N$ TTM-chains, from the perspectives of computational load and communication volume.

\section{Computational Load}
The TTM component performs $N$ TTM-chains, each involving $(N-1)$ TTM operations.
Commutativity allows us to rearrange and reuse the operations in multiple ways,
all of which can be represented in the form of TTM-trees, as observed in prior wok \cite{ucar-report}.
We measure the computational load of a tree by the number of floating point operations incurred.
Our objective is to design an efficient algorithm for finding the optimal TTM-trees.
Below, we first formalize the above model and rephrase prior schemes, and then describe the optimal algorithm.

\subsection{TTM-trees and Cost}
In a TTM-tree, the root represents the input tensor $\tenT$, 
each leaf node represents a unique new factor matrix and each internal node (nodes other than the root and the leaves) 
represents TTM along a particular mode. The root-to-leaf path leading to a new factor matrix $\newF_n$
realizes the TTM-chain required for computing $\newF_n$.

\subsubsection*{TTM-Trees}
Formally, a {\em TTM-tree} $H$ is a rooted tree with a function $\lbl(\cdot)$ that assigns
a label $\lbl(u)$ to each node $u$ such that the following properties are satisfied:
(i) the label of the root node is $\lbl(\rt) = \tenT$;
(ii) there are exactly $N$ leaves, with each leaf $u$ being labeled with a unique new factor matrix $\lbl(u)=\newF_n$;
(iii) each internal node $u$ is labeled with a mode $\lbl(u)\in [1,N]$;
(iv) for each leaf $u$ with label $\lbl(u)=\newF_n$, 
the path from the root to $u$ has exactly $(N-1)$ internal nodes and all the modes except $n$ appear on the path.

\begin{figure*}
\centering
\begin{tabular}{ccccc}
	\includegraphics[width=1.5in]{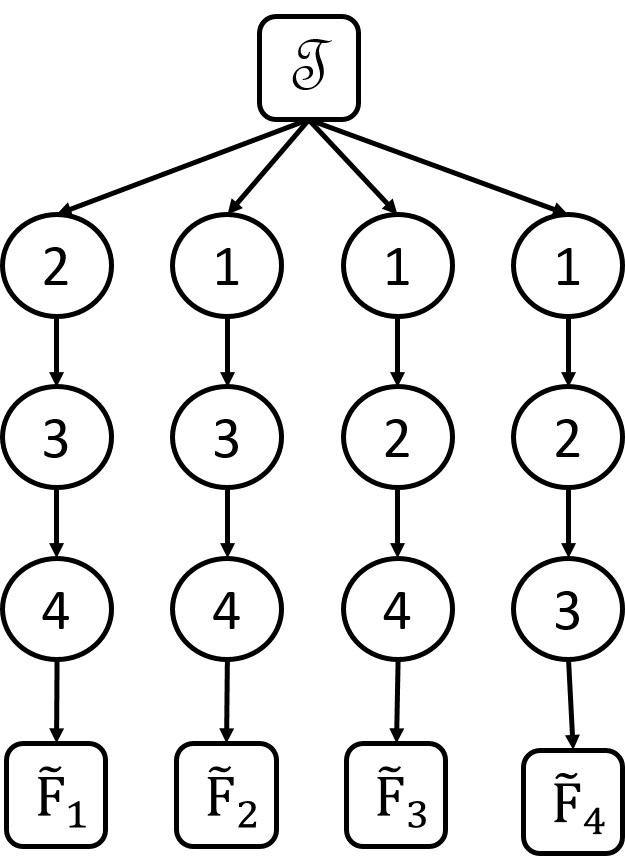} & \quad &
	\includegraphics[width=1.5in]{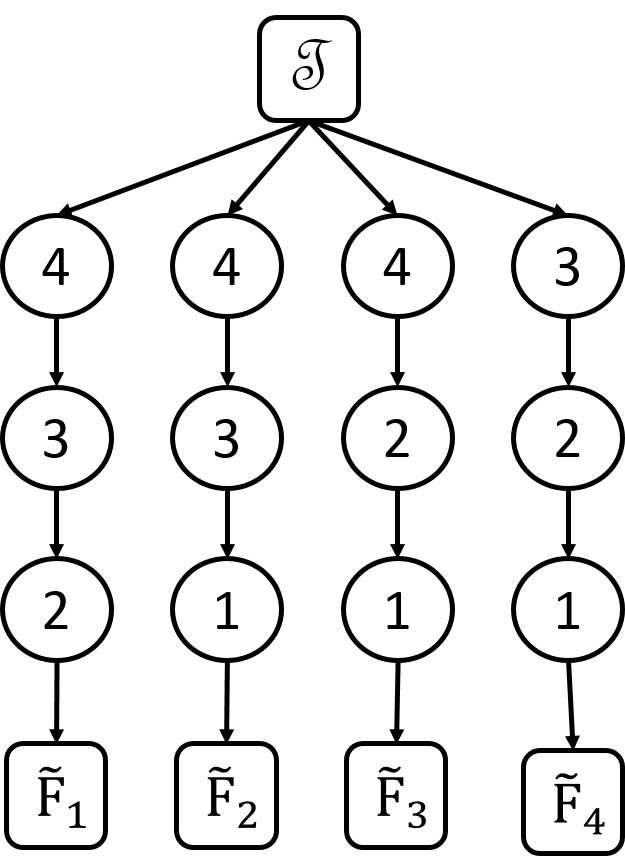} & \quad &
\includegraphics[width=1.5in]{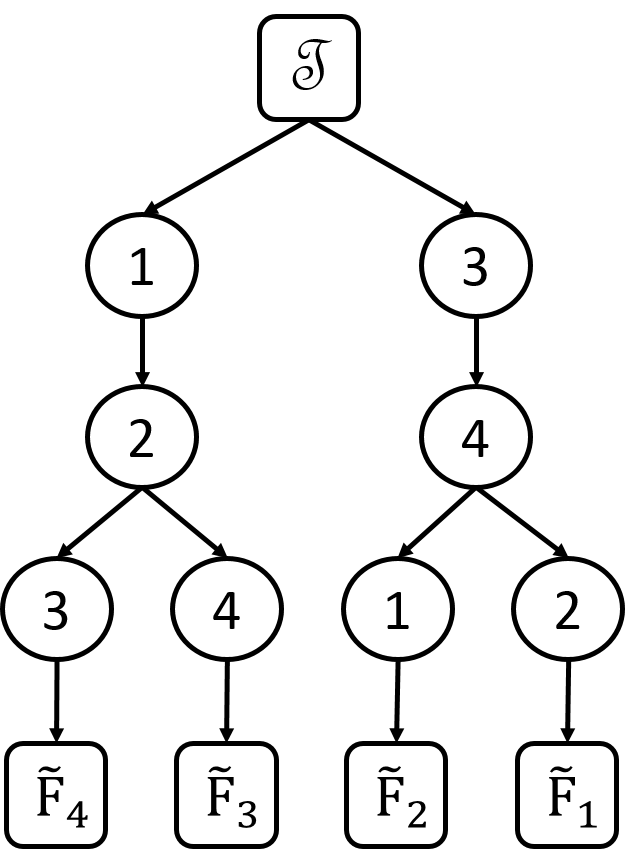} \\
(a) Chain tree & \quad & (b) Chain tree & \quad & (c) Balanced tree  
\end{tabular}
\caption{Example TTM-trees. Tree (a) and (b) are both chain trees, but use different orderings,
$\langle 1,2,3,4\rangle$ and $\langle 4,3,2,1\rangle$, respectively.}
\label{fig:load-ex}
\end{figure*}

Figure \ref{fig:load-ex} (a) - (c) provides example TTM-trees for the case of $N=4$.
Although the trees differ in the order in which the modes are processed and the total number of TTMs performed, 
they all realize the necessary TTM chains.

Given a tree $H$, the HOOI procedure can be executed via a natural top-down process
by associating each node with an input tensor $\In{u}$ and an output tensor $\Out{u}$.
For the root node, $\In{\rt} = \Out{\rt}= \tenT$. 
Each internal node $u$ with $\lbl(u)=n$ 
takes as input the tensor output by its parent $v$, multiplies it along mode $n$ by the factor matrix $\tmatF_n$,
and outputs the resultant tensor, i.e., $\In{u} = \Out{v}$ and $\Out{u} = \In{u}\times_n \tmatF_n$.
Each leaf node $u$ with $\lbl(u)=\newF_n$
constructs the new factor matrix $\newF_n$ by performing an SVD on the tensor output by its parent.
The correctness of the procedure follows from the commutativity property of the TTM-chain operation.
In the above procedure, we reuse the tensor output by a node for processing all its children.
By executing the process via an in-order traversal,  we can 
ensure that the maximum number of intermediate tensors stored at any point is bounded by the depth of the tree.

\subsubsection*{Computational Load}
We define the {\em cost (or computational load)} of a TTM-tree $H$ to be the number of floating point operations (FLOP) performed.
Each internal node $u$ with label $\lbl(u)=n$ executes the TTM $\Out{u} = \In{u}\times_n \tmatF_n$.
Recall that the operation involves the matrix-matrix multiplication, wherein the matrix $\tmatF_n$ is multiplied by the mode-$n$ unfolding of $\In{u}$.
The matrix has size $K_n \times L_n$ and the unfolded tensor has size $L_n \times (|\In{u}|/L_n)$
and so, the cost of the TTM is $K_n \cdot |\In{u}|$.  The cardinality of the output tensor is $|\Out{u}| = (K_n/L_n)|\In{u}|$;
namely, the node compresses the tensor by a factor $(K_n/L_n)$.
We can compute the cost incurred at all the nodes and the cardinality of their output tensors
by performing the above calculations in a top-down manner. 
Then, the cost of the tree $H$ is given by the sum of costs of its internal nodes. 
We can see that each mode $n$ is associated with two parameters: 
a {\em cost factor} $K_n$ and a {\em compression factor} $(K_n/L_n)$, which we denote as $h_n$.
At each node, the cost incurred and the cardinality of the output tensor can be expressed in terms of these two parameters.

\begin{figure*}
\centering
\includegraphics[width=5in]{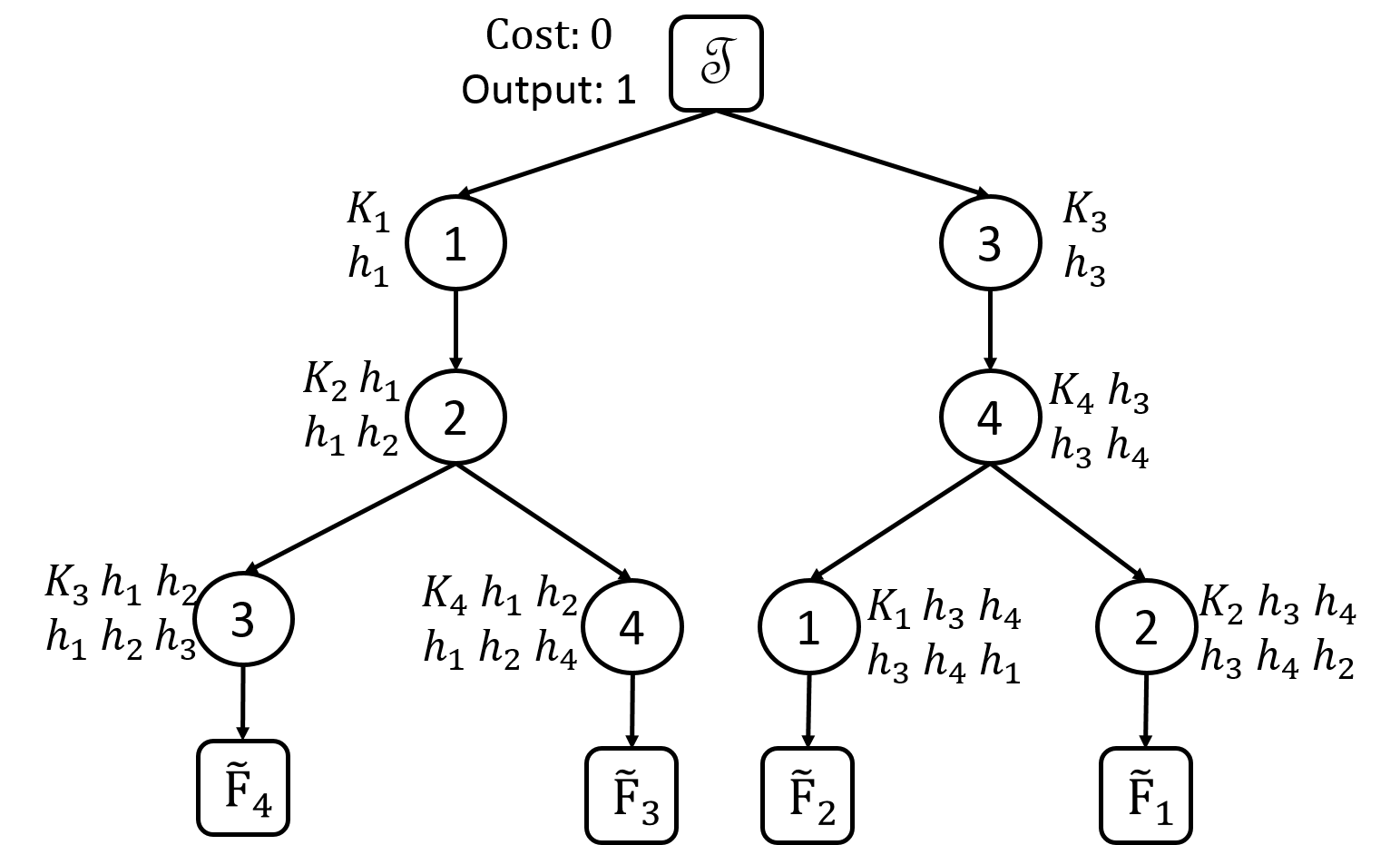} 
\caption{Cost analysis}
\label{fig:cost-ex}
\end{figure*}

Figure \ref{fig:cost-ex} provides an illustration. The cost incurred and the cardinality of the output tensor are shown at each node.
For the ease of exposition, we have normalized all the quantities by $|\tenT|$. 
The root node has cost $0$ and its cardinality of its output is $|\tenT|$, which is $1$ after normalization. 
Each node $u$ with label $n$ incurs a cost of $K_n$ times the cardinality of the tensor output by its parent;
it outputs a tensor having cardinality compressed by a factor $h_n$.

\subsection{Prior Schemes}
\label{sec:prior}
We rephrase the prior schemes in terms of TTM-trees. 

\paragraph{Chain trees}
These trees encode the naive scheme, with $N$ independent chains, each comprising of $(N-1)$ nodes (see figure \ref{fig:load-ex} (a) and (b)).

\paragraph{Balanced trees}
Chain trees perform $N(N-1)$ TTMs. 
Kaya and {\ucar} \cite{ucar-report} improved the count to approximately $N\log N$, via a divide-and-conquer strategy.
The idea is to divide the modes into two groups $\{1,2,\ldots, m\}$
and $\{m+1, m+2,\ldots, N\}$, where $m=\myfloor{N/2}$. 
We create a chain of nodes of length $m$ with labels from the first group and attach it to the root.
Then, we recursively construct a subtree for the second group and attach it at the bottom of the chain.
We then repeat the process by reversing the roles of the two groups.
Figure \ref{fig:load-ex} (c) shows an example for $N=4$. The number of internal nodes is approximately $N\log N$.

\paragraph{Mode Ordering}
Since the TTM-chain operation is commutative, the TTM products within a chain can be performed in any order.
Based on this fact, Austin et al. \cite{kolda-ipdps} propose the concept of mode ordering,
wherein the modes of the input tensor are rearranged according to some permutation.
For example, Figure \ref{fig:load-ex} (a) and (b) are both chain trees, but have different mode orderings.
They proposed two greedy heuristic for mode ordering.
The first heuristic arranges the modes in increasing order of cost factor $K_n$, placing lower cost modes at the top of the tree where large tensors are encountered. 
The second heuristic arranges the modes in increasing order of compression factor $h_n$, aiming at higher compression at the top layers of the tree.
We are not aware of any prior work on mode ordering with respect to balanced trees. 

\subsection{Constructing Optimal Trees}
In this section, we present our algorithm for constructing the optimal TTM-tree, the tree with the minimum cost.
The algorithm is based on dynamic programming and runs in time $O(4^N)$. 
In practice, the algorithm takes negligible time, 
since the number of dimensions of dense tensors is fairly small (typically, $N\leq 10$).

Towards developing the dynamic programming algorithm, we first claim that the optimal TTM-tree is binary,
namely every node has at most two children. 
The proof is based on the observation that 
if a node $u$ has three children, then the children can be rearranged so that
only two of the nodes remain as children of $u$. 
We then identify a set of subproblems and derive a recurrence relation relating them.
This is followed by a description of the algorithm and an analysis of the running time.

\begin{figure}
\centerline{
	\includegraphics[width=5in]{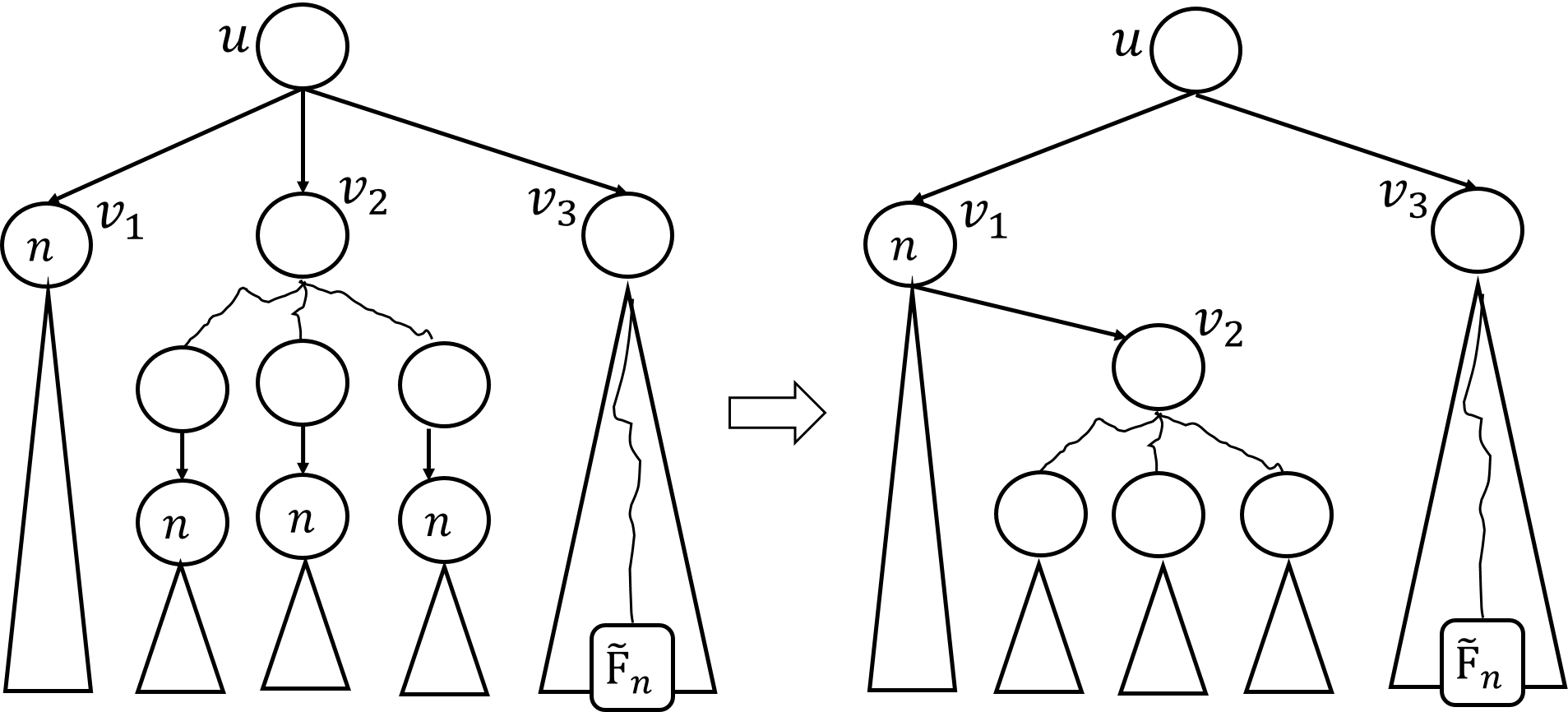}
}
\caption{Illutration of binary tree transformation}
\label{fig:bintree}
\end{figure}

\begin{lemma}
\label{lem:binary}
There exists an optimal binary tree.
\end{lemma}
\proof
Let $H^*$ be an optimal tree. Suppose a node $u$ has three children $v_1$, $v_2$ and $v_3$.
The properties of TTM-trees imply that none of the three nodes can be a leaf node.
Let $H_1$, $H_2$ and $H_3$ be the subtrees rooted at the three nodes
and let $n$ be the label of $v_1$. Without loss of generality, assume that the leaf node bearing the label $\newF_n$ appears in the subtree $H_3$.
Below, we argue that $H^*$ can be transformed in to a new tree $H'$, without increasing the cost, 
such that $v_2$ is no longer a child of $u$. Thus, the transformation reduces the number of children of $u$ by one.
By repeating the process, we can get a binary tree having cost not more than $H^*$.
The transformation is discussed next.

If the label of $v_2$ is also $n$, then we can merge $v_1$ and $v_2$. 
Otherwise, mode $n$ must appear on all the paths from $v_2$ to the leaves in $H_2$. 
We perform two operations: (i) for any node $z$ in $H_2$ with label $n$, we delete the node (by making its children the children of its parent);
(ii) make $v_2$ as a child of $v_1$. Let $H'$ be the new tree created by the process.
See Figure \ref{fig:bintree} for an illustration.

We can verify that $H'$ is a valid TTM-tree. Furthermore, the cost of $H'$ cannot be more than $H^*$, as argued next.
The cost of the nodes outside $H_2$ does not change. Let $z'$ be any node in $H_2$ and consider three cases:
(i) if $z'$ is one of the deleted nodes, then we save its cost;
(ii) if $z'$ is the descendant of a deleted node, its cost does not change;
(iii) if $z'$ is an ancestor of a deleted node, the cost cannot increase,
since under $H'$, the tensor input to $z'$ is further shrunk by TTM along mode $n$.
\qed

\subsubsection*{Subproblems}
Consider any binary tree $H$ and let $u$ be an internal node in it. 
With respect to $u$, the modes $n$ can be partitioned into three groups:
(i) {\em pre-multiplied}: $n$ is found along the path from the root to $u$, including $u$;
(ii) {\em computed under $u$}: the leaf bearing label $\newF_n$ is found under the sub-tree rooted at $u$;
(iii) $n$ does not belong to either category.
Let $P$, $Q$ and $R$ denote the set of modes belonging to the three categories. 
For an illustration, consider the tree in Figure \ref{fig:load-ex} (c) and let $u$ denote the right child of the root labeled $3$;
with respect this node, $P=\{3\}$, $Q=\{1,2\}$ and $R=\{4\}$.
Notice that the triple $(P,Q,R)$ forms a partitioning of $[1,N]$.
We can characterize any node $u$ in a TTM-tree via the above $3$-partition. 

We next make an observation regarding the set $R$. 
Consider the stage in the HOOI execution, wherein we have completed the processing of the node $u$.
At this stage, we have already completed multiplication along all modes in $P$.
For any mode $n\in Q$, the corresponding TTM-chain involves multiplication along all modes, except $n$.
Of these modes, we are yet to perform multiplication along the modes in $R$ and $Q\setminus \{n\}$.
The multiplications along modes in $R$ are common to the TTM-chains corresponding to all the modes in $Q$.
Therefore, at this stage, we can potentially select any mode from $R$, 
perform multiplication along the mode and reuse the output tensor. 
Hence, we call the modes in $R$ as {\em reusable}.  
For instance, mode $4$ is reusable in the example discussed earlier (Figure \ref{fig:load-ex}).

The idea behind the dynamic programming algorithm is to consider a subproblem for each possible triple $(P,Q,R)$
as follows: construct the optimal subtree given that the modes in $P$ have been multiplied already,
the modes in $Q$ needs to be computed and $R$ are the reusable modes. 
We formalize the concept using the notion of partial TTM-trees.
These trees are similar to the usual TTM-trees, 
except that the root represents a partially processed tensor and we only need compute a partial set of factor matrices.

\paragraph{Partial TTM-tree} 
Consider a triple $(P,Q,R)$ with $|Q|\geq 1$. Let $\ttmcP$ denote the tensor obtained by multiplying $\tenT$
by the factor matrices along all the modes found in $P$.
A {\em partial TTM-tree} for $(P,Q,R)$ is a rooted tree with labels on its nodes
such that the following properties are satisfied:
(i) the root is labeled $\tenX =\ttmcP$;
(ii) there are exactly $|Q|$ leaves, with each leaf $u$ being labeled with a unique factor matrix $\newF_n$, for $n\in Q$;
(iii) each internal node $u$ is labeled with a mode from $[1,N]\setminus P$;
(iv) for each leaf node $u$ with label $\newF_n$, the path from the root to $u$
has exactly $N-|P|-1$ internal nodes and all the modes except $P\cup \{n\}$ appear on them.
Figure \ref{fig:counter} shows two example partial-TTM trees for the triple
$P=\{3\}$, $Q=\{1,2\}$ and $R=\{4\}$ discussed earlier.

The cost of a partial-TTM tree is defined analogous to the usual TTM-trees.
Let $H^*(P, Q, R)$ denote the optimal partial TTM-tree for the triple $(P,Q,R)$
and let $\optcost(P,Q,R)$ be the cost of the optimal tree.
The optimal tree for the original problem
is given by $H^*(P,Q,R)$ with $P=\emptyset$, $Q=[1,N]$ and $R=\emptyset$.

\subsubsection*{Recurrence Relation}
We discuss the subproblem structure and derive a recurrence relation.
Consider a triple $(P,Q,R)$.  Since optimal trees are binary, the root of $H^*(P,Q,R)$ can have either one or two children.
The recurrence relation considers both the possibilities, which we refer to as {\em reuse} and {\em splitting}.

{\it Reuse: }
This option is available, if $R\neq \emptyset$.
In this case, we select a mode $n\in R$ and multiply $\tenX=\ttmcP$ along mode $n$. 
The result is then reused for computing the new factor matrices of all the modes in $Q$.
In terms of TTM-trees, the operation corresponds to adding a single child with label $n$ to the root of the partial TTM-tree.
Once the above TTM operation is performed, 
we are left with solving the subproblem corresponding to the triple $(P\cup \{n\}, Q, R\setminus\{n\})$.
The cost is given by sum of the cost of the TTM operation $\tenX \times_n \tmatF_n$
and the cost of recursively solving the subproblem. Recall that the former cost is $K_n\cdot |\tenX|$.
The latter cost is $\optcost(P\cup \{n\}, Q, R\setminus\{n\})$.
In the above process, any mode from $R$ can be reused and 
we can find the best option by considering all the choices.

{\it Splitting:}
The second possibility is to split (or partition) $Q$ into sets $Q_1$ and $Q_2$
and independently solve the triples $(P, Q_1, R)$ and $(P, Q_2, R)$.
The total cost is given by the sum of costs of optimal subtrees of the two subproblems,
i.e., $\optcost(P, Q_1, R) + \optcost(P, Q_2, R)$.
Any (non-trivial) partition $(Q_1,Q_2)$ of $Q$ with $Q_1,Q_2\neq \emptyset$
can be used in the above process and the best choice can be found by an exhaustive search.

The above discussion yields the following recurrence relation for computing 
the optimal cost of a triple $(P,Q,R)$:
\[
	\optcost(P,Q,R) = \min\{\optcost_{\rm reuse}, \optcost_{\rm split}\}, \mbox{where}
\]
\begin{eqnarray*}
	\optcost_{\rm reuse} &=& \min_{\substack{n\in R}} K_n\cdot |\ttmcP| + \optcost(P\cup \{n\},Q,R\setminus\{n\})\\
	\optcost_{\rm split} &=& \min_{\substack{\langle Q_1, Q_2\rangle \subseteq Q}} \optcost(P,Q_1,R) + \optcost(P,Q_2,R).
\end{eqnarray*}

\subsubsection*{Algorithm and Running Time Analysis}
The algorithm constructs a dynamic programming table having at most $3^N$ entries, one for each triple $(P,Q,R)$ with $|Q|\geq 1$.
The base cases for the recurrence relation are triples with $|P|=N-1$, $|Q|=1$ and $|R|=0$, and the cost is $0$ in these cases.
The other entries get computed by looking up previously computed entries as per the recurrence relation.
The entries can be considered according to the following partial ordering:
$(P',Q',R')$ precedes $(P,Q,R)$, if either $P\subset P'$ or $P=P'$ and $Q'\subset Q$.
The optimal partial trees can be constructed in a similar manner. 
The optimal tree for the original problem corresponds to the triple $P=\emptyset$, $Q=[1,N]$ and $R=\emptyset$.

The running time of the algorithm can be analyzed by counting the number of dynamic programming table lookups performed.
Each lookup can be specified by a configuration of the form $\langle P\cup\{n\}, Q, R-\{n\},n\rangle$ in the reuse scenario,
and by $\langle P,Q_1, Q_2,R\rangle$ in the splitting scenario.
In either case, there are at most $4^N$ possible configurations.
The algorithm does not perform lookup on the same configuration twice
and hence, the total number of lookups is at most $2\cdot 4^N$.  
Thus, algorithm runs in time $O(4^N$).

\begin{figure}
\centerline{
\includegraphics[width=2in]{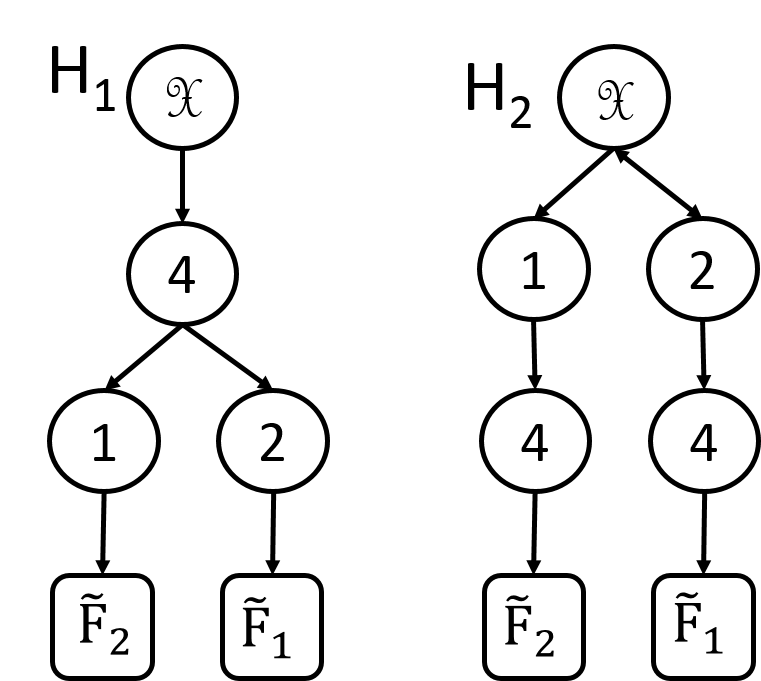}
}
\caption{Example partial TTM-trees with $N=4$, $P=\{3\}$, $Q=\{1,2\}$, and $R=\{4\}$. $\tenX=\tenT[P]=\tenT\times_3 \tmatF_3$}
\label{fig:counter}
\end{figure}

{\bf Remarks:}
In the recurrence relation, we may intuitively think that whenever $R\neq \emptyset$, we should always reuse some mode from $R$;
see tree $H_2$ in Figure \ref{fig:counter} for an illustration not reusing even though $R\neq \emptyset$.
However, the strategy is incorrect. We can construct examples, wherein thn optimal tree 
sacrifices the reuse option on modes having high cost factor so as to postpone multiplication along these modes till the tensor shrinks sufficiently

Given that $N$ is small, we may consider constructing the optimal TTM-trees via an exhaustive search.
A naive search over all TTM-trees is prohibitively expensive. 
The TTM operation corresponding to a mode $n$ involves multiplication along all the other $(N-1)$ modes,
which can be performed in any of the $((N-1)!)$ orderings. Over all the nodes, the number of combinations is $((N-1)!)^N)$, 
all which can be realized as chain trees. We can expedite the search by considering only the binary TTM-trees.
We are not aware of any closed form expression for the number of binary TTM-trees.
We note that our algorithm can be modified to enumerate all these trees.
Instead of enumeration, the algorithm incorporates memoization and computes the optimal tree efficiently in time $O(4^N)$.

\section{Communication Volume}
\label{sec:distribution}
Our strategy is to fix a TTM-tree $H$ (based on the heuristic or the optimal tree)
and devise schemes for minimizing the volume.
Our distributed implementation uses the same strategy as that of Austin et al. \cite{kolda-ipdps}
for distributing the tensors and performing TTM in a distributed manner.
We propose a dynamic gridding scheme that offers significant reduction in volume
and design an efficient algorithm for finding the optimal scheme.
%We first present a brief outline of the distributed setup.

\subsection{Distributed Setup}
\label{sec:setup}
\begin{figure}
\centering
\includegraphics[width=4.0in]{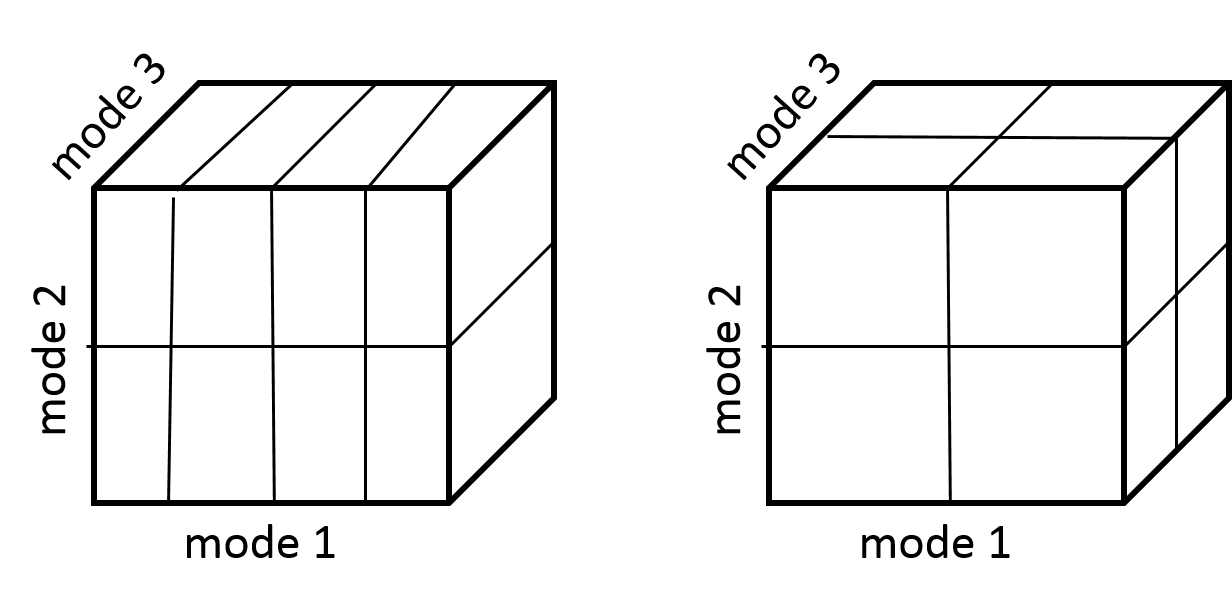}
\caption{Example grids: the two figures use the grids
$\langle 4,2,1\rangle$ and $\langle 2,2,2\rangle$, respectively.}
\label{fig:setup-a}
\end{figure}

\paragraph{Tensor Distribution}
Fix a TTM-tree $H$ and let $P$ be the number of processors. 
We arrange the processors in an $N$-dimensional grid $g = q_1 \times q_2 \times \cdots \times q_N$ such that $P=\prod_j q_j$.
To distribute a tensor, we impose the grid on the tensor and partition it into $P$ blocks,
and assign each block to a processor; see Figure \ref{fig:setup-a} for an illustration.
The input tensor $\tenT$ and all the intermediate tensors gets partitioned using the same grid.  

\paragraph{Distributed TTM and Volume}
Each node $u$ with label $n$ and parent $v$ performs the TTM operation $\Out{u} = \In{u}\times_n \tmatF_n$.
For the grid $g$, we denote the communication volume incurred by the operation as $\vol(u, g)$.
As observed in the prior work $\vol(u,g) = (q_n -1)|\Out{u}|$; a brief outline of the argument in the following paragraph.
The total communication volume of $g$, denoted $\vol(H,g)$, is defined to be the sum of volumes incurred at all the internal nodes.

\begin{figure}
\centering
\includegraphics[width=6in]{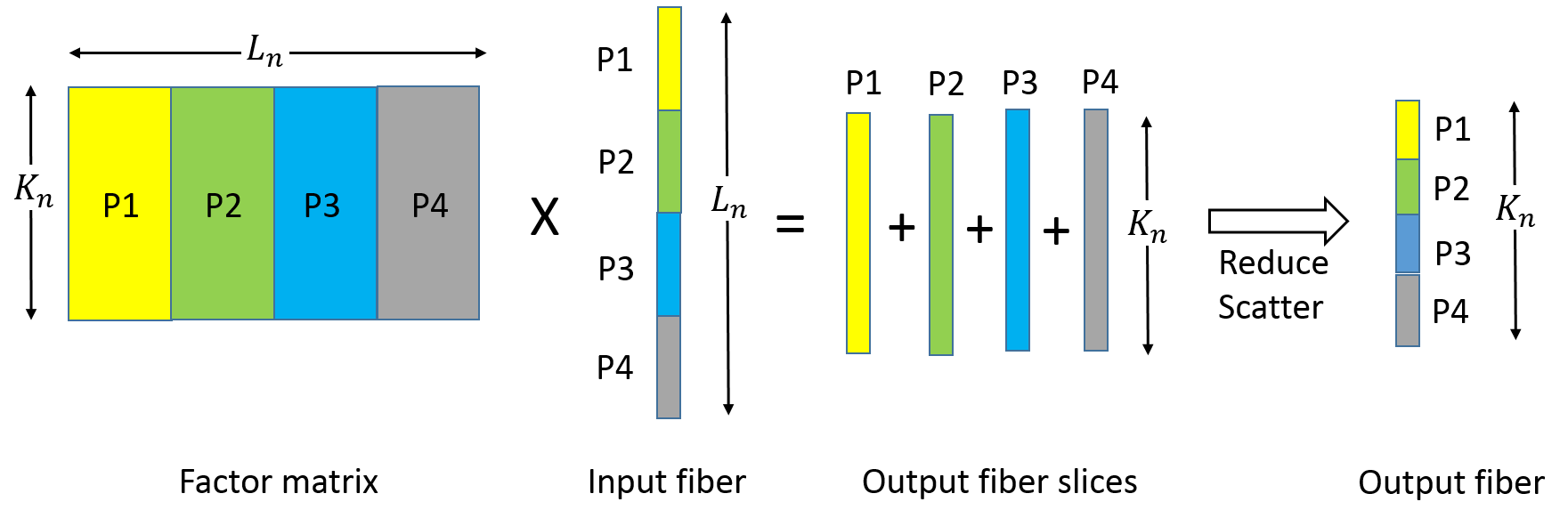}
\caption{Matrix-fiber multiplication}
\label{fig:setup-b}
\end{figure}

Recall that the TTM operation $\Out{u}=\In{u}\times_n \tmatF_n$ can be viewed as applying the linear transformation $\tmatF_n$
to every mode-$n$ fiber $\vect{x}$ of $\In{u}$. That is, we need to perform 
the matrix-vector product $\vect{y}=\tmatF_n\cdot \vect{x}$.
Since the factor matrices are small in size, we can afford to keep a copy of them at every processor. 
However, each mode-$n$ fiber $\vect{x}$ gets distributed equally among some $q_n$ processors 
and so, computing the product requires a reduce operation.
Similarly, the output fiber $\vect{y}$ must be distributed among the same processors using a scatter operation.
See Figure \ref{fig:setup-b} for an illustration.
The reduce-scatter operation is performed over the output fiber $\vect{y}$ of $K_n$,
for which we incur $(q_n-1)K_n$ units of communication. 
Summed up over all the fibers, the total communication volume for the TTM is $(q_n-1)|\Out{u}|$.

In the above distribution method, if $q_n > L_n$ for some mode $n$, then some processor would receive an empty block while partitioning $\tenT$.
Similarly, if $q_n > K_n$ then same scenario would arise on some intermediate tensor.
We avoid the load imbalance by considering only grids with $q_n\leq K_n$, for all $n$; we call these {\em valid grids}.
In the rest of the discussion, unless explicitly mentioned, we shall only consider valid grids.

\begin{table}[]
\centering
\begin{tabular}{l|llllll}
          & $N =$ 5 & 6     & 7    & 8     & 9     & 10    \\ \hline
$P = 2^5$ & 126     & 252   & 562  & 792   & 1287  & 2002  \\
$2^{10}$    & 1001    & 3003  & 8008 & 19448 & 43758 & 92378 \\
$2^{20}$    & 10626   & 53130 & 230K & 880K  & 3.1M  & 10M  
\end{tabular}
\caption{Number of grids for differnt values of $P$ and $N$}
\label{fig:numgrids}
\end{table}

\subsection{Finding the Optimal Static Grid}
\label{sec:psi}
We observe that the optimal static grid, the one achieving the minimum communication volume, can be found via an exhaustive search in negligible time.
The number of grids, including the invalid ones, is the same as number of ways in which the integer $P$ can be expressed as the product of $N$ factors, which we denote $\psi(P, N)$.
If the prime factorization of $P$ is $P = p_1^{e_1}\cdot p_2^{e_2}\cdots p_s^{e_s}$, then we have that
\[
	\psi(P, N) = \prod_{i=1}^s {e_i + N - 1 \choose N-1}
\]
Table \ref{fig:numgrids} shows the quantity for example values of $P$ and $N$. 
When the quantity becomes large, the search can be parallelized in a straightforward manner.
Even for the extreme case of $P=2^{20}$ and $N=10$, the number of grids to be scanned per processor is approximately $10$.

\subsection{Dynamic Gridding Scheme}
%The gridding scheme discussed in the previous section
%uses a single grid for representing all the intermediate tensors in the tree,
%which we refer to as {\em static gridding}.
%In this section, we propose a dynamic gridding scheme that offers
%significant reduction in communication volume, even when compared to the optimal statics grids.

The idea of dynamic gridding is as follows. Consider a node $u$, and let its parent be $v$ and label be $n$. 
The node $u$ performs the TTM operation $\Out{u}=\In{u}\times_n \tmatF_n$.
If the tensor $\In{u}$ is represented in a grid $g=\langle q_1, q_2, \ldots, q_N\rangle$ then we incur a volume of $(q_n-1)|\Out{u}|$,
Thus, it is beneficial to represent $\In{u}$ under a grid with a small assignment $q_n$,
and in fact, the operation can be made communication-free by assigning $q_n =1$.
The static gridding scheme selects a single grid by considering the cumulative effect of the above communication volume over all the nodes.
The idea of dynamic gridding is to select different grids for representing the intermediate tensors,
as appropriate for each node.
However, we need to pay a price for dynamic gridding: 
if the tensor output by the parent $v$ is represented in a grid $g$
and we have selected a different grid $g'$ for representing it at $u$,
then the tensor must be regridded (redistributed) among the processors.
The process incurs a volume of $|\In{u}|$. Thus, a dynamic grid scheme must decide whether or not to regrid at each node,
and furthermore, if it decides to regrid, the new grid must be selected in a manner
beneficial for the TTM operations performed later in the subtree,
so that the overall communication is minimized.

\begin{figure}
\centerline{
\includegraphics[width=5in]{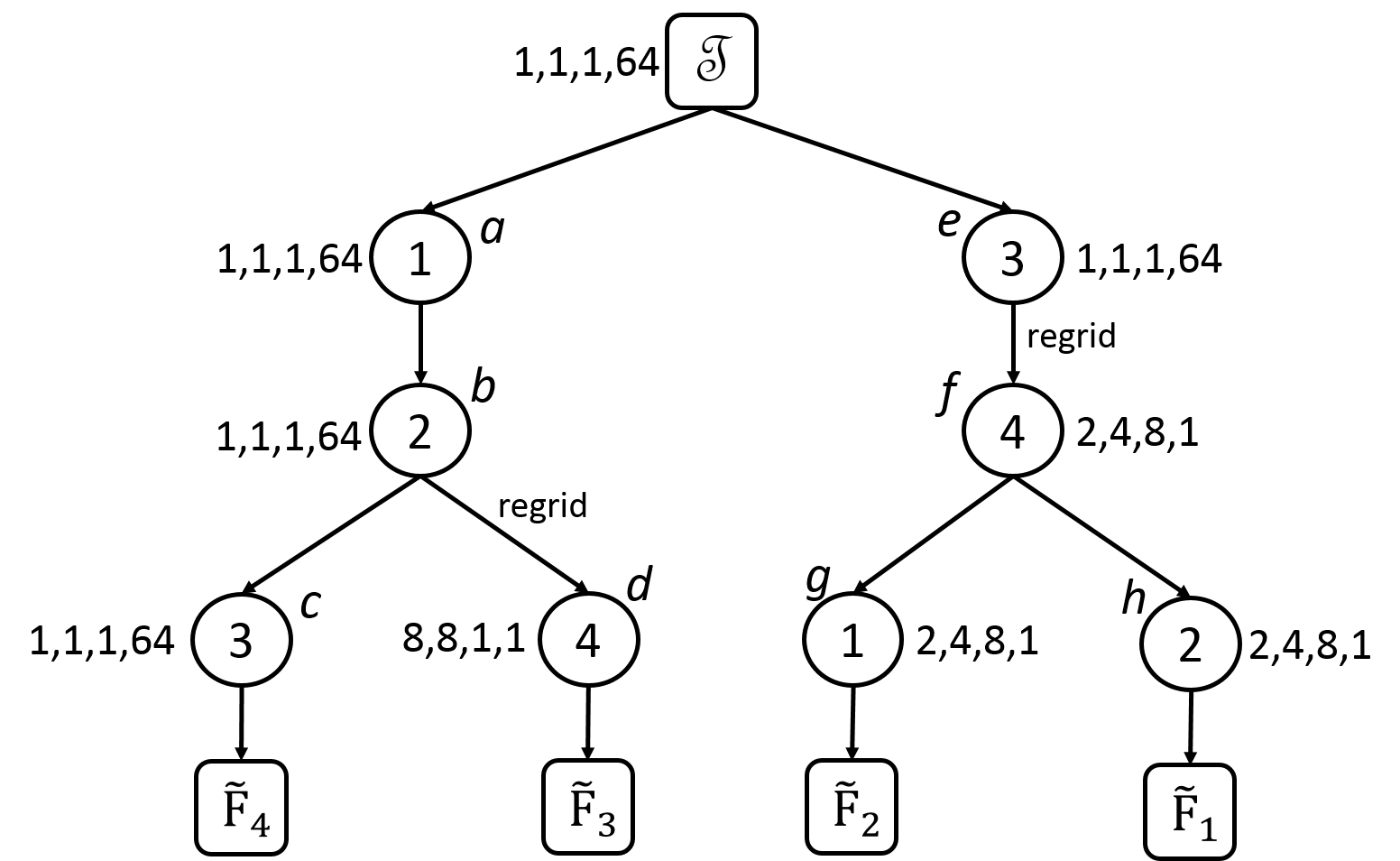}
}
\caption{Example dynamic grid scheme}
\label{fig:dyna}
\end{figure}

In Figure \ref{fig:dyna}, we have shown an example, carefully constructed so as to highlight the different aspects of dynamic gridding. 
Assume that the number of processors is $P=64$ and the core is of size $8 \times 8\times 8\times 64$. 
The choice of the initial grid $\langle 1,1,1,64\rangle$ 
makes the TTM operations at nodes $a,b,c$ and $e$ are communication-free. 
However, the grid is not suitable for the TTM at node $d$, since the volume incurred is $63 \times |\Out{d}|$.
Instead, we switch to a new grid $\langle 8,8,1,1\rangle$, making the operation communication-free.
We perform another regrid operation at node $f$ by selecting the new grid $\langle 2,4,8,1\rangle$,
The choice of the new grid is motivated by the following considerations. 
The subtree beneath $f$ does not involve any TTM along mode $3$ and so, it is prudent to assign a high value along the mode.
However, we must select a valid grid, and the constraint implies that 
the maximum possible value is $8$ (since the core length along mode $3$ is $K_3 = 8$).
We next assign a value of $1$ to mode $4$, thereby making the TTM at node $d$ communication-free.
The remaining of value of $8$ is assigned to the modes $1$ and $2$ in a balanced manner.

\paragraph{Dynamic Grid Scheme} 
Formally, a {\em dynamic grid scheme} is a mapping $\pi$ that associates a grid $\pi(u)$ with each node $u$. 
The volume incurred by the scheme, denoted $\dvol(H,\pi)$ is defined as follows.
For each node $u$ with label $n$ and parent $v$, we compute the volume incurred at the node as the sum of two components:
(i) TTM operation volume: $(q_n-1)|\Out{u}|$, where $q_n$ is the assignment to mode $n$ under $\pi(u)$;
(ii) regridding volume: if $\pi(u)$ is the same as the parent grid $\pi(v)$, 
then the volume is zero, and otherwise, it is $|\In{u}|$.
The volume of the scheme $\pi$, denoted $\dvol(H, \pi)$, is defined to be the sum of communication incurred over all the nodes $u$.
At the root node, we represent the input tensor $\tenT$ under the grid $\pi(\rt)$ and we do not have the regrid option.
Let $\dvol^*(H)$ denote the optimal communication volume achievable among all dynamic grid schemes,
and let $\opt(H)$ denote an optimal scheme.

\subsection{Optimal Dynamic Gridding Scheme}
In this section, we develop an efficient dynamic programming algorithm for computing the optimal dynamic grid scheme for a given tree $H$.
For a node $u$, let $\subT{u}$ denote the subtree rooted at $u$. 
A {\em partial grid scheme} for $\subT{u}$ refers to a mapping $\pi$ that specifies a grid for each node in $\subT{u}$.
For each node $u$ and each grid $\gpar$, we shall define a subproblem with the following connotation:
assuming that the tensor output by the parent is represented under the grid $\gpar$,
find the optimal partial grid scheme for the subtree $\subT{u}$.
We solve these subproblems via a bottom-up traversal of the tree,
wherein the optimal solution at $u$ is computed from the optimal solutions of its children.

\subsubsection*{Subproblems}
Consider a pair $(u, \gpar)$, where $u$ is a node and $\gpar$ is a grid.
For a partial grid scheme $\pi$ for the subtree, let $\dvol(\subT{u},\pi|\gpar)$ denote
the volume incurred by $\pi$ given that the tensor output by the parent of $u$ is represented in the grid $\gpar$.
Formally, it is computed as follows.  For each node $z\in \subT{u}$, define a parent grid $\pg(z)$: 
for the node $u$, $\pg(u) = \gpar$, and for the other nodes, $\pg(z)=\pi(z')$, where $z'$ is the parent of $z$.
For any node $z\in \subT{u}$, associate the volume given by the sum of the following two components:
(i) TTM operation volume: $(q_n-1)|\Out{z}|$, where $n$ is the mode label of $z$ and $q_n$ is the assignment to mode $n$ under $\pi(z)$;
(ii) regridding volume: if $\pi(u)$ is the same as $\pg(z)$, then the volume is zero, and otherwise, it is $|\In{z}|$.
Then, the volume $\dvol(H_u, \pi|\gpar)$ is defined to be the sum of volumes associated with all the nodes $z\in \subT{u}$. 
Let $\dvol^*(\subT{u}|\gpar)$ denote the minimum volume possible among all partial grid schemes $\pi$. 
We do not regrid at root and so, define $\dvol^*(H|\gpar)$ to be the minimum volume given that $\tenT$ is represented under $\gpar$. 

\subsubsection*{Recurrence Relation}
We derive a recurrence relation for computing $\dvol^*(\subT{u}|\gpar)$. 
Let $v_1, v_2, \ldots, v_s$ be the children of $u$. 
In determining the optimal partial scheme, we have two options:
(i) regrid: select a new grid $\rg(u)$ for representing $\In{u}$;
(ii) do no regrid: represent $\In{u}$ under the given grid $\gpar$.
In the first case, we select $\rg(u)$ to be the grid yielding the minimum volume for the child subtrees:
\[
	\rg(u) = \argmin_{\substack{g}} \quad \sum_{j=1}^s \dvol^*(\subT{v_j}|g).
\]

We can now write the recurrence for $\dvol^*(\subT{u}|\gpar)$. Let $n$ be the label of $u$ and $v$ be the parent of $u$.
Let $\gpar = \langle p_1, p_2, \ldots, p_N\rangle$ and let $\rg(u) = \langle q_1, q_2, \ldots, q_N\rangle$. Then:
\[
\dvol^*(\subT{u}|\gpar) = \min\{\vol^*_1, \vol^*_2\},\mbox{where}
\]
\[
\vol^*_1 = |\In{u}| + (q_n-1)|\Out{u}| + \sum_{j=1}^s \dvol^*(\subT{v_j}|\rg(u))
\]
\[
\vol^*_2 = (p_n-1)|\Out{u}| + \sum_{j=1}^s \dvol^*(\subT{v_j}|\gpar)
\]
The two quantities correspond to the optimal solutions for the two choices of regridding and not regridding.
In both the cases, we incur communication for the TTM operation and communication in the subtrees.
In addition, the first case incurs a regrid volume of $|\In{u}|$.
Under the two choices, the tensors $\In{u}$ and $\Out{u}$ get represented under the grids $\rg(u)$ and $\gpar$, respectively.
Consequently, the recursive calls for the two choices are made with the corresponding grids.
At the root node, we represent $\tenT$ under $\gpar$ and do not regrid
and so, we consider only the first choice at the root.
The optimal volume for the whole tree $\dvol^*(H)$ is given by minimum of $\dvol^*(H|\gpar)$, over all the choices of $\gpar$
and can be computed via enumerating the choices.

\subsubsection*{Algorithm and Running Time Analysis}
The algorithm constructs a dynamic programming table containing an entry for each pair $(u,\gpar)$.
Thus, the number of entries is $|H|\cdot \psi(P, N)$, where $|H|$ is number of nodes in the tree.
The entries are computed via a bottom-up traversal of the tree.
For each entry $(u, \gpar)$, we need to compute $\rg(u)$, which requires a search over all the grids. 
However, the selection of the grid $\rg(u)$ is independent of the parameter $\gpar$ and so, it is sufficient to compute it once per node.
The recurrence relation involves a table lookup for each child and an entry for a node is looked up only by its parent,
and so the total number of table lookups is $O(|H|\cdot \psi(P,N))$.
Similar to the case of static grids (Section \ref{sec:psi}), 
the exhaustive search involved in computing $\rg(u)$ can be parallelized in a straightforward manner, if $P$ is large.
Thus, the algorithm executes in negligible time in practice.

\section{Distributed Implementation}
\label{distr_impl}
The distributed implementation consists of two modules, a {\em planner} and an {\em engine}.
The planner constructs a TTM-tree, either based on the heuristics or the optimal tree, 
and selects grids, either the optimal static or dynamic gridding scheme.
The module only requires the meta-data as input: the dimension lengths of the input tensor $\tenT$ and  the core tensor.
It needs to be executed only once and the output can be used across multiple invocations of the HOOI procedure.
All the processors use the same TTM-tree and there is synchronization at each tree node. 

The second module, called the {\em engine} maintains tensors in a distributed manner, and implements the TTM and SVD routines.
Tensors are distributed according to the block distribution method (Section \ref{sec:distribution}).
To change the grid under which a tensor is represented, the engine implements an element redistribution procedure via the {\tt MPI\_Alltoallv} collective. 
The TTM operation is implemented in a distributed manner using the algorithm proposed by Austin et al. \cite{kolda-ipdps}. 
The naive method for computing the TTM product along a specified mode first requires an unfolding of the tensor along the mode.
Their algorithm cleverly avoids the unfolding operation by employing a blocking strategy
which breaks down the TTM product into a series of matrix multiplication calls. The matrix multiplication calls are performed using {\tt dgemm}. 
We implement the SVD component using distributed Gram matrix computation ($AA^T$) followed by eigen value decomposition (EVD).  
The Gram matrix computation is performed using {\tt dysrk} calls which exploits the symmetry in the product. 
The EVD is computed sequentially by invoking the {\tt dsyevx} routine;
this is acceptable since it operates on small square matrices of size $L_n\times L_n$,
and $L_n \leq 2000$ in our setting.

\newcommand{\pltscale}{0.45}
\newcommand{\bigplt}{0.45}
\section{Experimental Evaluation}
In this section, we present an experimental evaluation of the algorithms described in the paper.

\subsection{Setup}
\subsubsection*{System}
The experiments were conducted on an IBM BG/Q system.
Each BG/Q node has $16$ cores and $16$ GB memory.
Our implementation is based on MPI and OpenMP, with {\tt gcc 4.4.6} and {\tt ESSL 5.1}. 
Each MPI rank was mapped to a single node and spawns 16 threads which are mapped to the cores. 
All the experiements use $32$ nodes.

\subsubsection*{Tensors}
As discussed in the introduction, the execution time of the HOOI algorithm is crucially dependent on the metadata (dimension lengths of the input tensor and the core tensor), 
and independent of the elements in the tensor. We exploit this property to construct a large benchmark of tensors with metadata derived 
from real world tensors considered in prior work. 

We also include a set of tensors with metadata derived from simulations in combustion science \cite{kolda-ipdps}.
The metadata of these tensors is shown in Table \ref{tab:real_tensor_dims}. Due to memory limitations, we curtailed the length along certain dimensions;
while the length along all the spatial dimensions were retained as
such, we reduced the length along the axes of variables/timesteps and proportionately reduced the length of the core along these axes.
We fill these tensors with randomly generated data. 

The benchmark is constructed as follows. 
%The input to the HOOI procedure consists of a tensor $\tenT$ of size $L_1\times L_2\times \cdots \times L_N$ and the desired core tensor size of $K_1, K_2, \ldots, K_N$. 
We constructed $5$ and $6$-dimensional tensors with dimension lengths $L_n$ drawn from the set $\{20, 50, 100, 400\}$. We selected the core dimension lengths $K_n$ by fixing the compression ratio $h_n=K_n/L_n$. The value for $h_n$ was drawn from the set $\{1.25, 2, 5, 10\}$. Given the above two sets of choices, an input for the HOOI procedure can be generated as follows: 
for each dimension $n\in [1,$N$]$, we select $L_n$ from the first set of choices, and select $h_n$ from the second set of choices, and set $K_n = h_n\cdot L_n$.
We placed an upper limit of $8\cdot 10^9$ on the cardinality of $\tenT$.
We enumerated all possible HOOI inputs in the above manner and obtained a benchmark consisting of $1134$ $5$-dimensional and $642$ $6$-dimensional tensors.

\begin{table}
\centering
\begin{tabular}{l|l|l}
     & Tensor Dimensions       & Core Tensor Dimensions \\ \hline
HCCI & (672, 672, 627, 16)     & (279, 279, 153, 14)    \\
TJLR & (460, 700, 360, 16, 4)  & (306, 232, 239, 16, 4) \\
SP   & (500, 500, 500, 11, 10) & (81, 129, 127, 7, 6)  
\end{tabular}
\caption{Real tensors used in our study}
\label{tab:real_tensor_dims}
\end{table}

\begin{figure*}[t!]
\centering
\subfloat[Overall time (5D)]{ \includegraphics[scale=\bigplt]{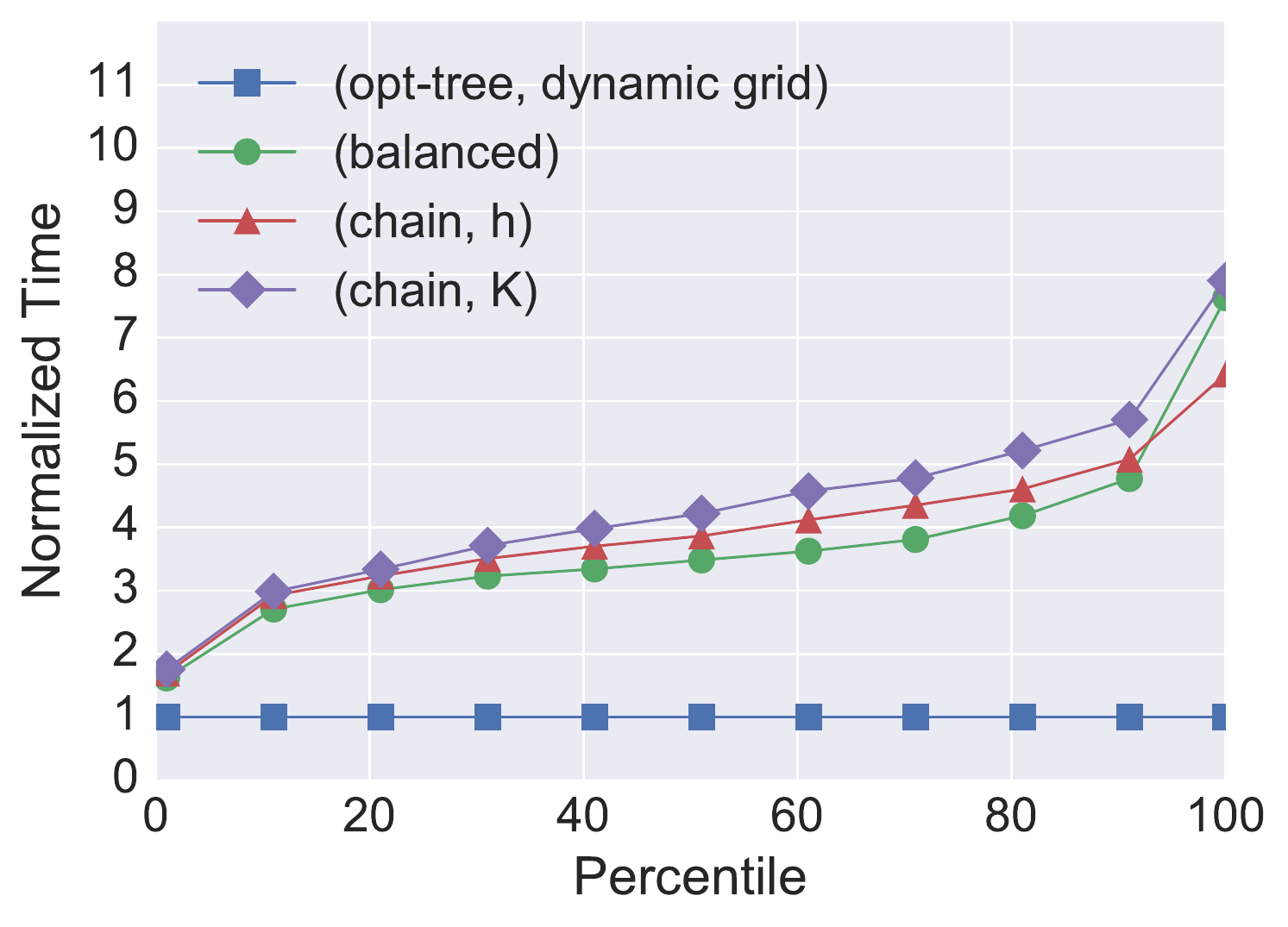} \label{fig:overall_time_5D}} \quad\quad
\subfloat[Overall time (6D)]{\includegraphics[scale=\bigplt]{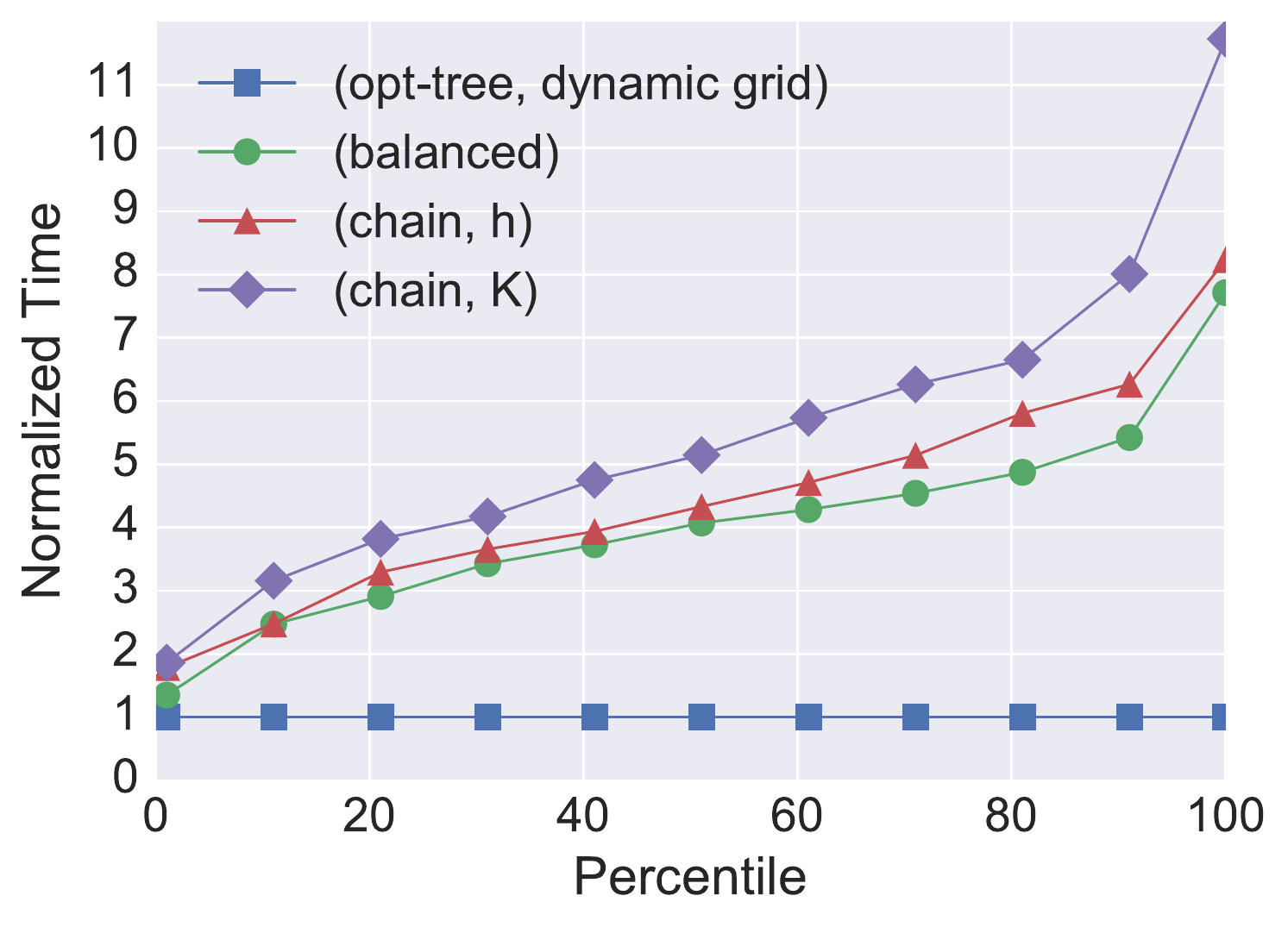} \label{fig:overall_time_6D}} \\
\subfloat[Real Tensors. CK:(chain,$K$), CH: (chain, $h$), B: (balanced), OPT:(opt-tree, dynamic grid)]{\includegraphics[scale=0.7]{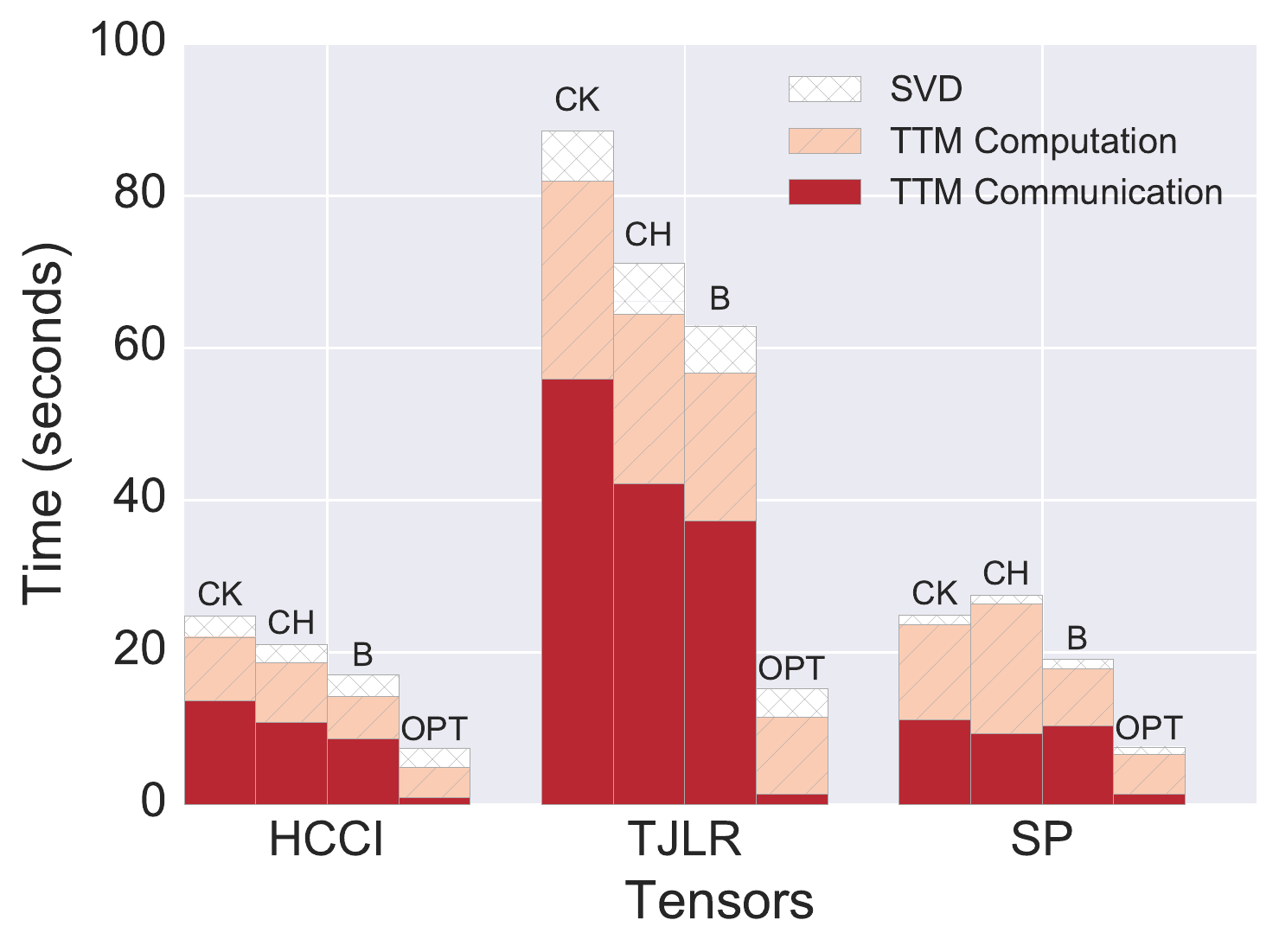} \label{fig:real_improvements}}
\caption{Overall Execution Time}
\end{figure*}

\begin{figure*}
\centering
\subfloat[Computational Time (5D)]{ \includegraphics[scale=\pltscale]{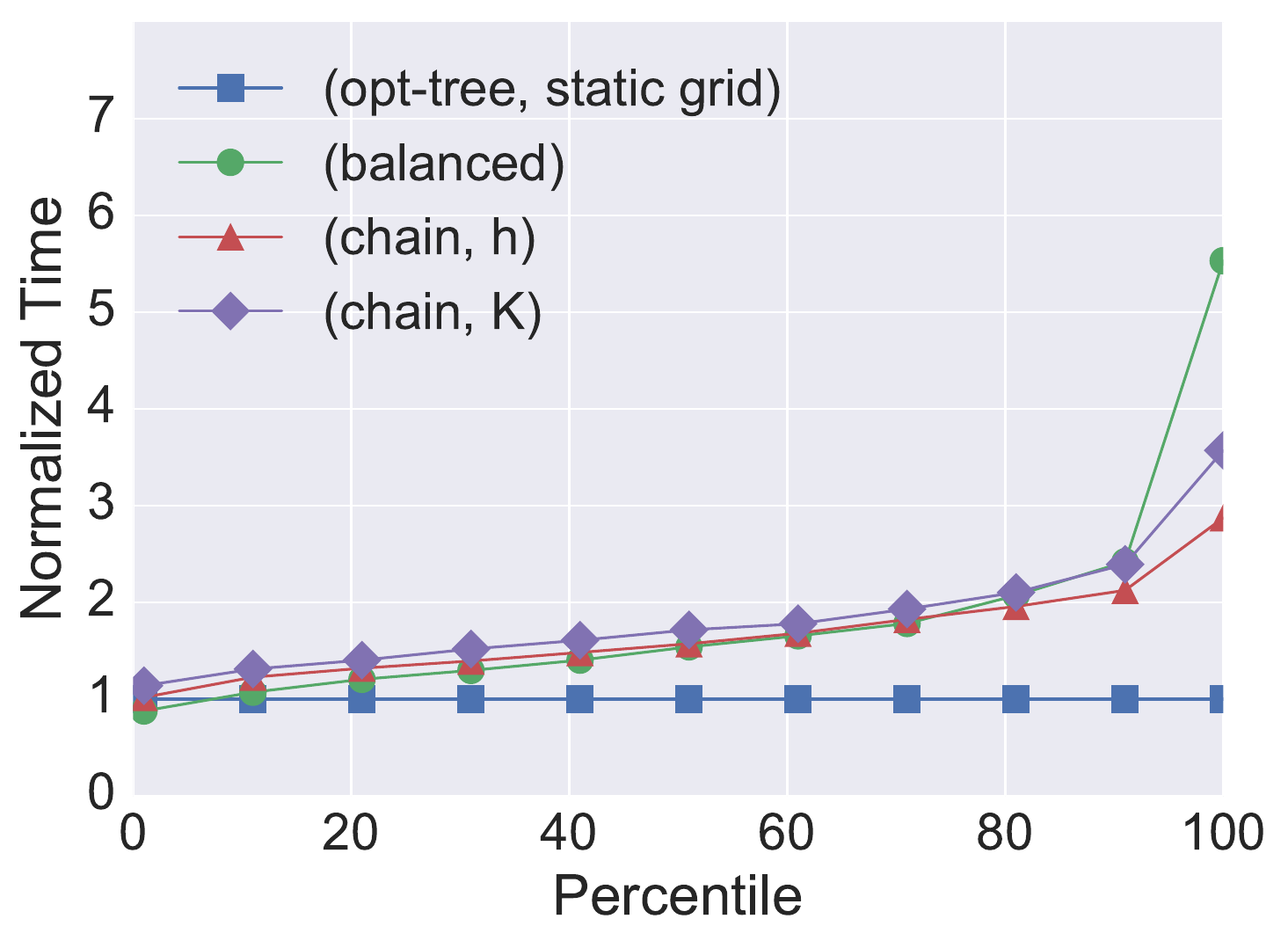} \label{fig:ttm_comp_time_5D}}\quad\quad
\subfloat[Computational Time (6D)]{\includegraphics[scale=\pltscale]{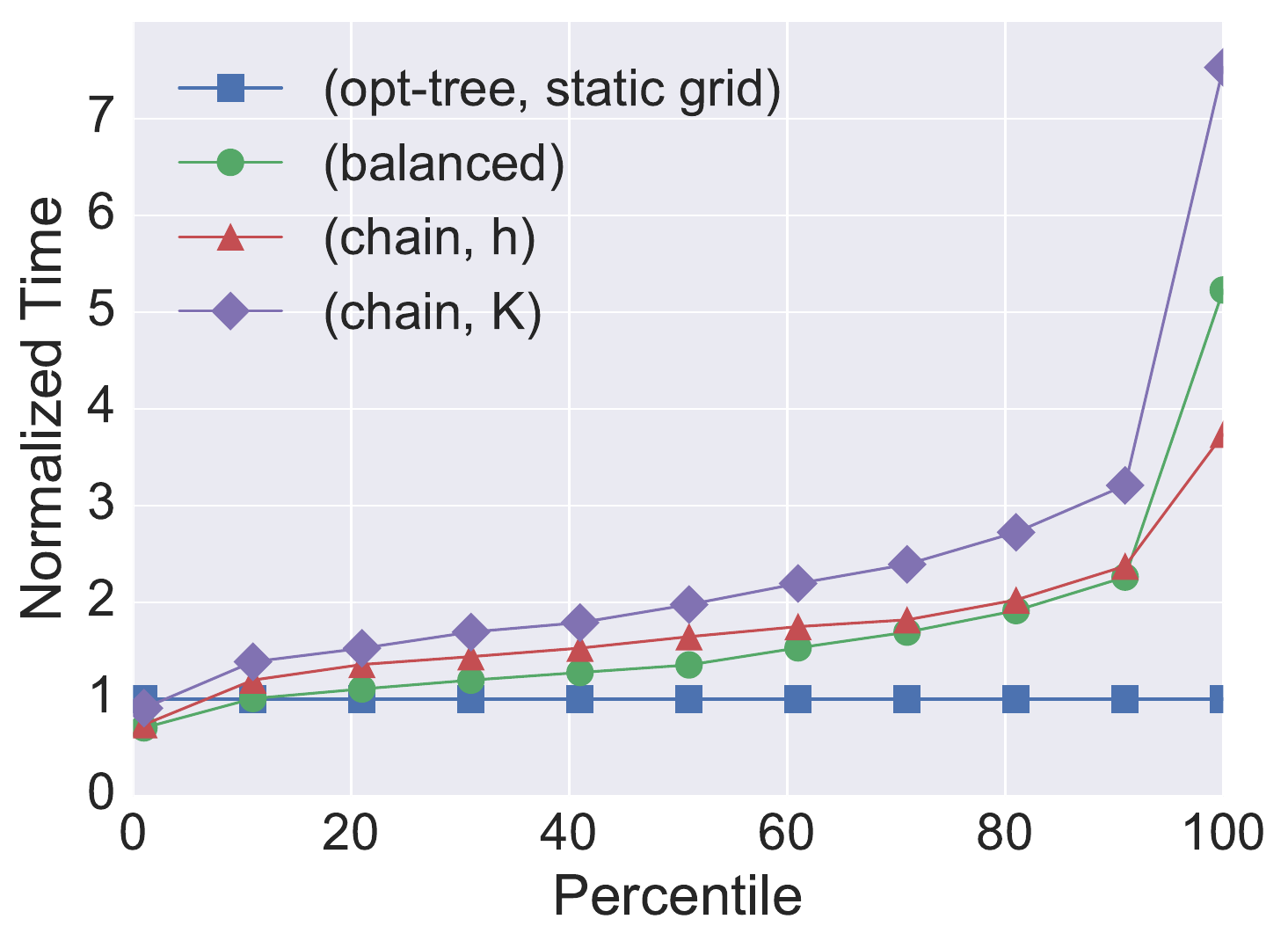} \label{fig:ttm_comp_time_6D}} \\
\subfloat[Computational Load (5D)]{\includegraphics[scale=\pltscale]{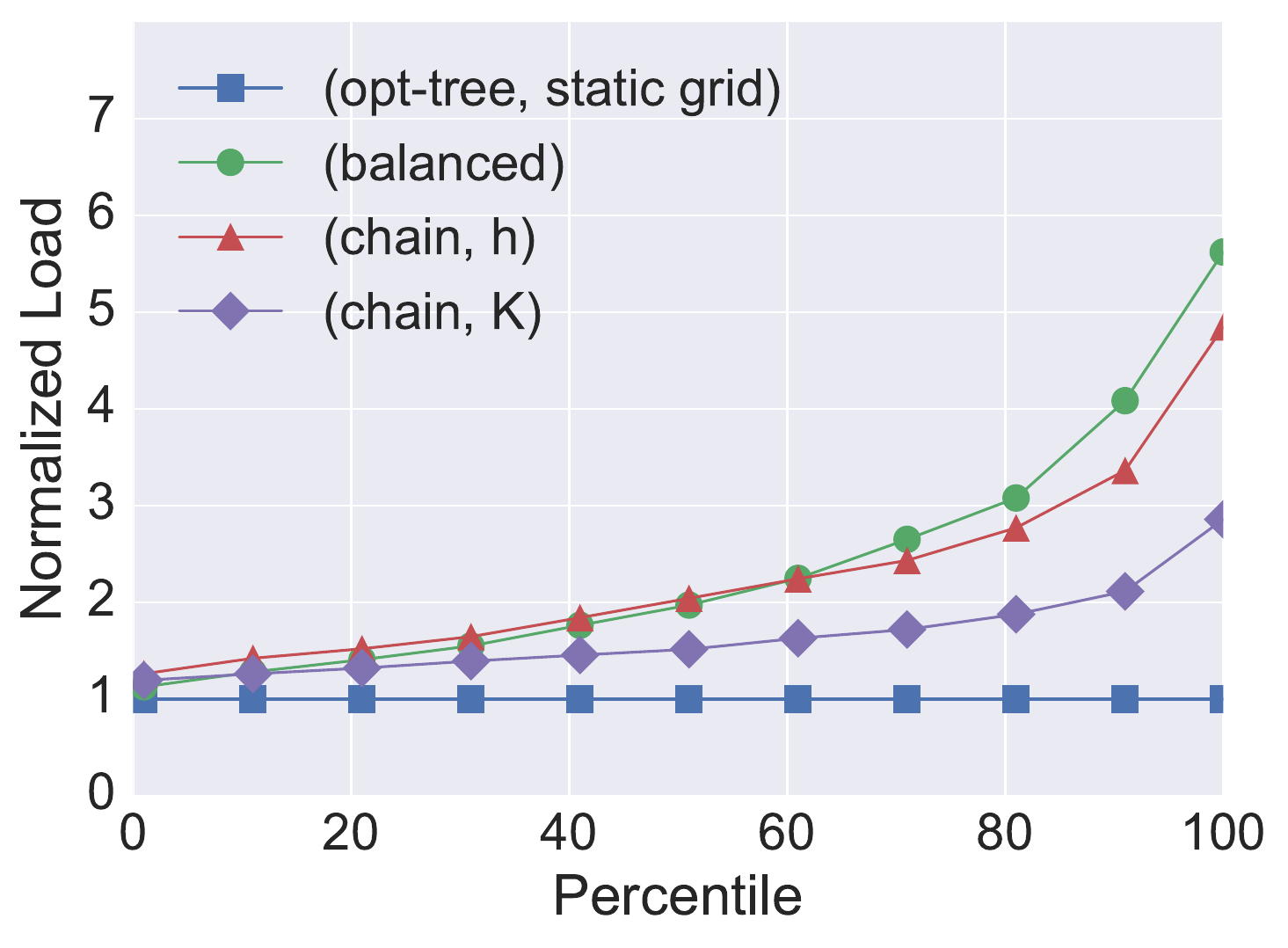} \label{fig:ttm_comp_load_5D}} \quad \quad
\subfloat[Computational Load (6D)]{\includegraphics[scale=\pltscale]{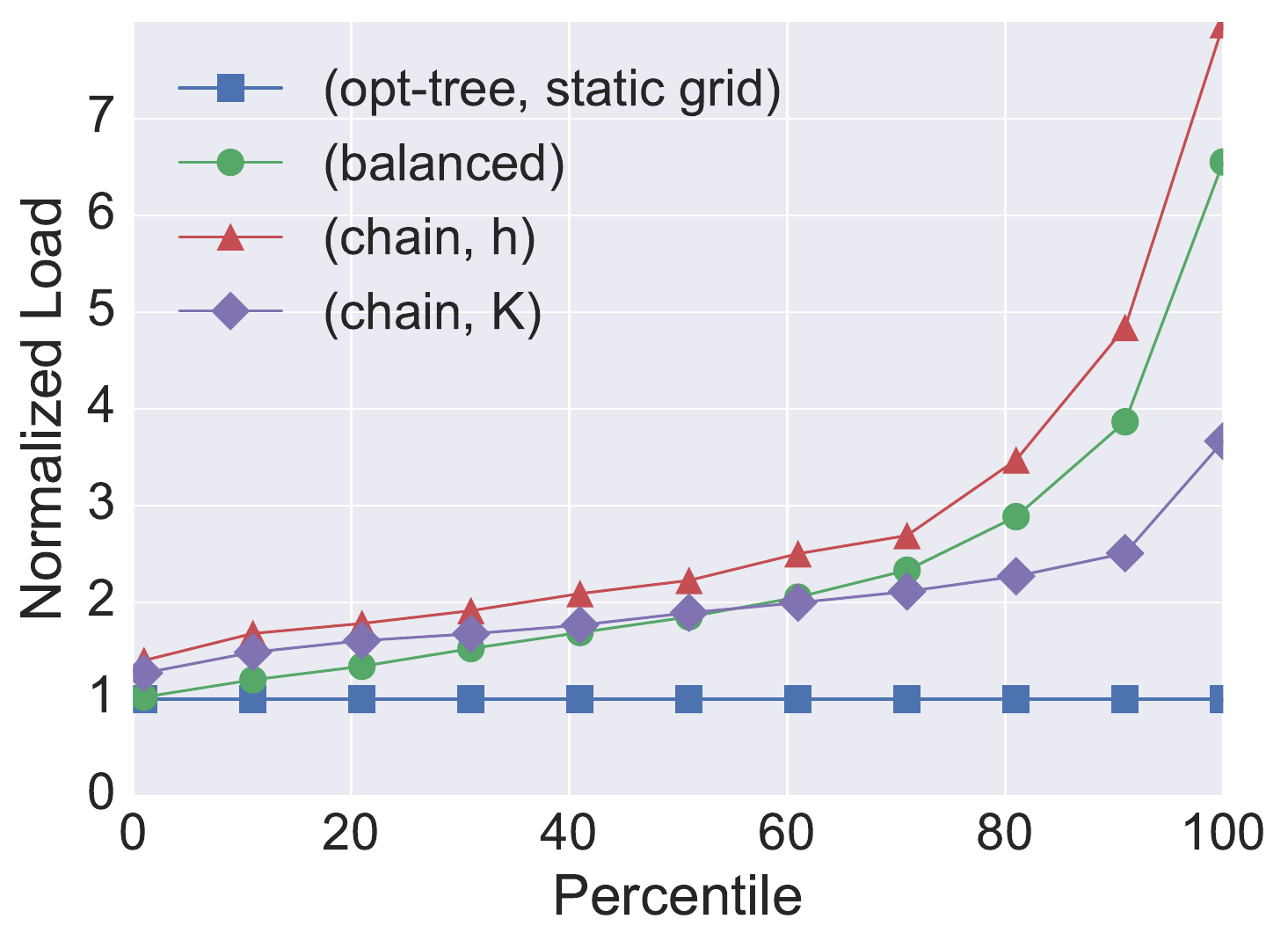} \label{fig:ttm_comp_load_6D}}\\
\subfloat[Commnunication Time]{ \includegraphics[scale=\pltscale]{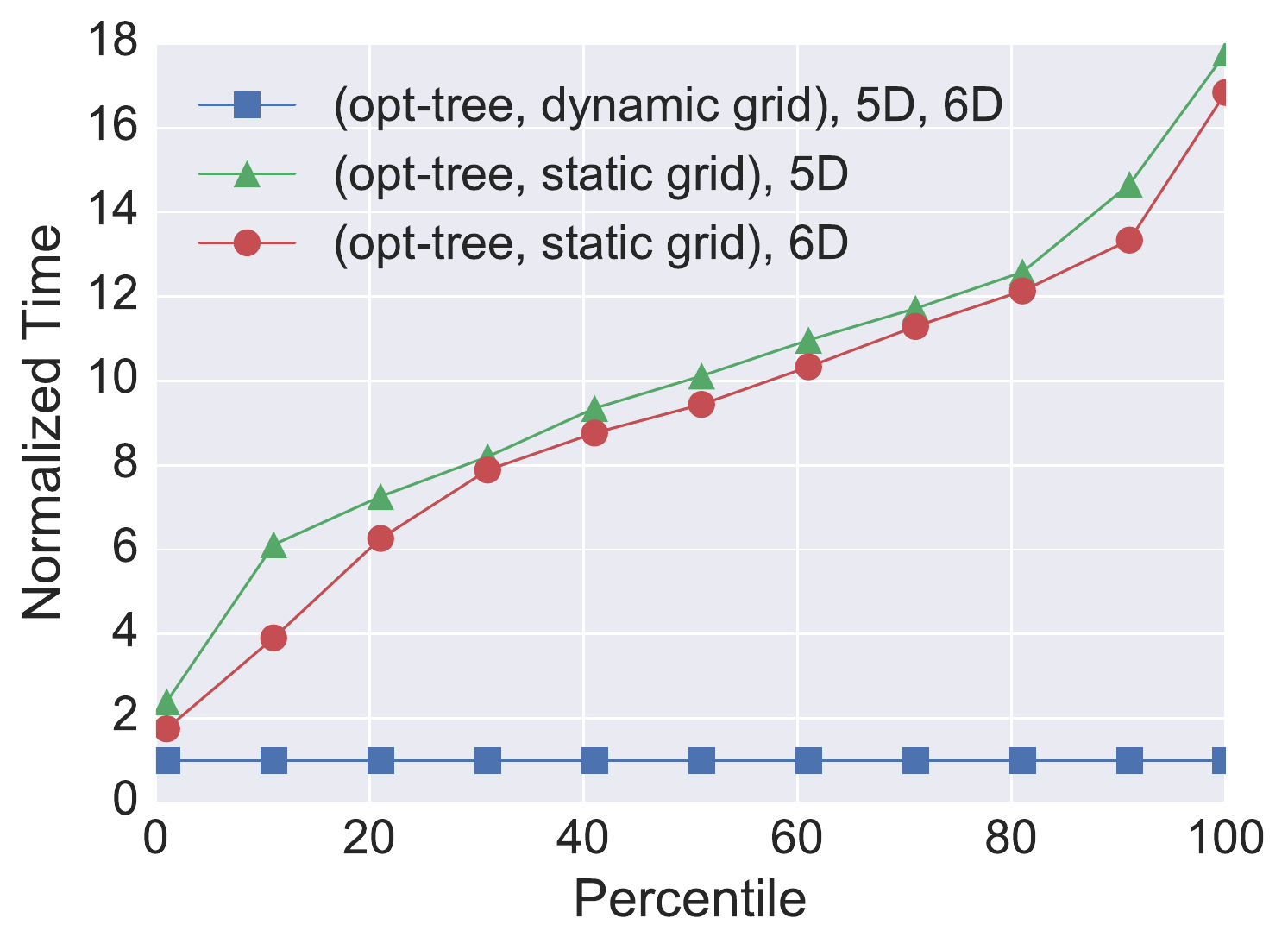} \label{fig:comm_time}}\quad \quad
\subfloat[Communication Volume]{\includegraphics[scale=\pltscale]{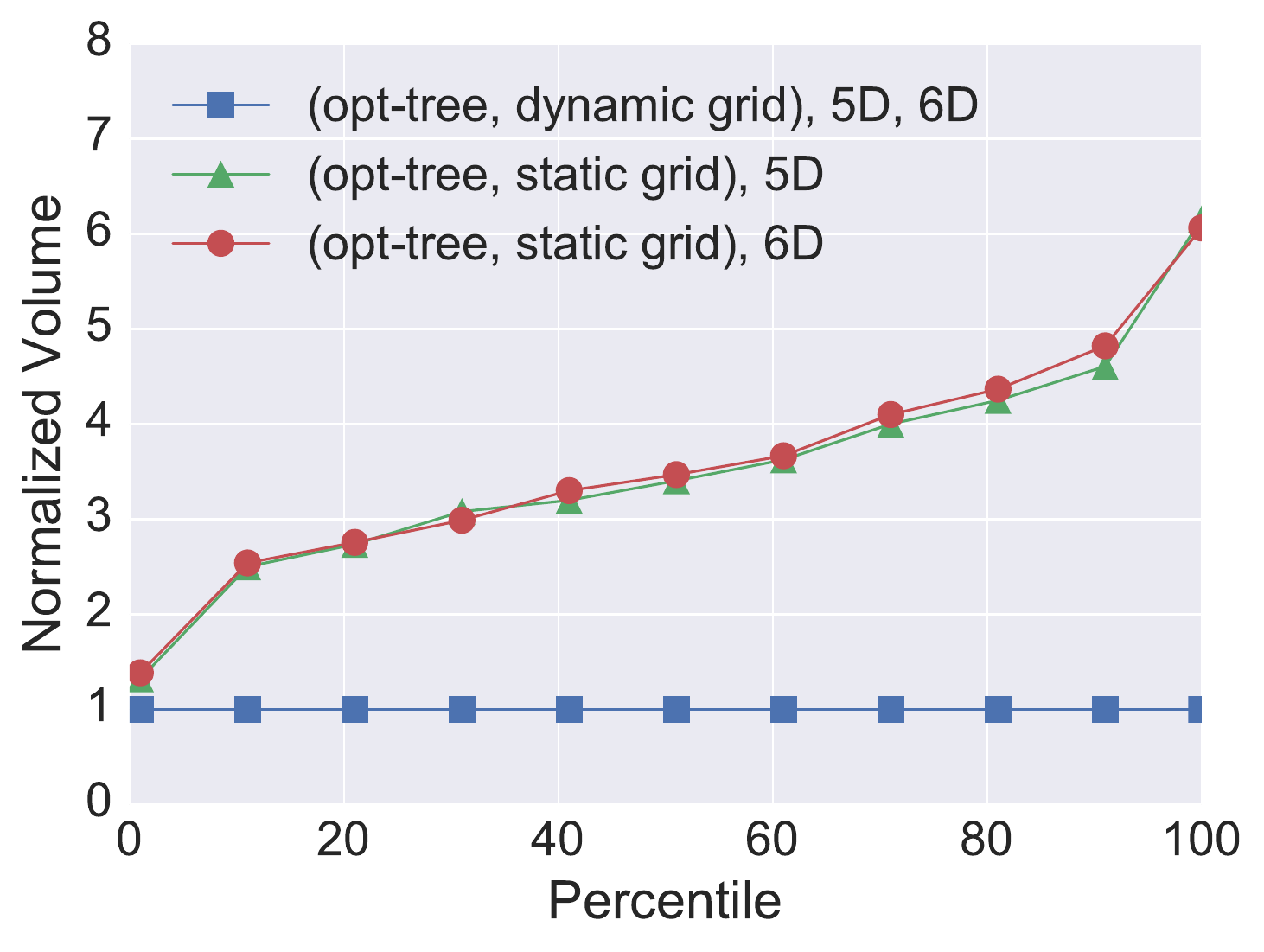} \label{fig:comm_load}}\\

\caption{Analysis of Benchmark Results}
\end{figure*}

\subsection{Evaluation}
The experiments involved comparing our algorithm and prior heuristics (Section \ref{sec:prior} and \ref{sec:setup}).
The heuristics are obtained by fixing the tree class to be chain and balanced trees,
and the mode ordering to be $K$-ordering and $h$-ordering. In the case of balanced trees, 
we observed that $K$-ordering and $h$-ordering do not impact the execution time and so, we use the input (naive) mode ordering.
For all these heuristics, we use the optimal static grids.
We compare the heuristics with our algorithms: the optimal tree algorithm with static grids and the same algorithm with dynamic grids.

The following metrics were studied: overall execution time, computational load and time,  and communication volume and time.
The dimensions of the tensors/matrices arising in the computations are identical across different HOOI iterations (only data elements change). 
Consequently, any two HOOI iterations will incur the same computational load and communication volume. Thus, the running times would be approximately the same across iterations.
Hence, we executed each algorithm on all the benchmark tensors and measured these metrics for a single HOOI invocation.

% In this section, we present the results on the $5$D and $6$D benchmarks using 32 nodes of BG/Q.
% Our results are summarized as follows:
% \begin{itemize}
% \item Overall improvements in HOOI iteration time are as high as 7x, compared to prior work (Section \ref{sec:expt_overall}).
% \item Optimal computation trees offer up to 3.7x reduction in computation time, compared to prior work (Section \ref{sec:expt_computation_opt}).
% \item Dynamic grids offer offer up to 17x improvement in communication time, compared to static grids (Section \ref{sec:expt_communication_opt}).
% \end{itemize}

\subsubsection*{Overall  Execution Time}
\label{sec:expt_overall}

%Median:
%3.4-4.2x over 5D
%4.0-5.1x over 6D
%Max:
%6.4-7.9x over 5D
%7.7-11.7x over 6D
We compared overall execution time of the opt-tree algorithm with dynamic gridding against the prior heuristics.
For each tensor, we normalized the execution times w.r.t the execution time of the opt-tree algorithm
(which becomes $1$ unit).
Given that the benchmark is large, we summarize the results using a percentile plot.
Figure \ref{fig:overall_time_5D} and \ref{fig:overall_time_6D}  shows the plots for $5$D and $6$D tensors.
In these plots, normalized time of $t$ on percentile value $k$ means that for $k\%$ of tensors, 
the normalized execution time is less than $t$. 
For example, in Figure \ref{fig:overall_time_5D}, the $60$th percentile value for the (chain, K) is $4.7$, 
meaning that the improvement factor obtained by the opt-tree algorithm is at most $4.7$x for $60\%$ of the tensors 
and at least $4.7$x for the remaining $40\%$ of the tensors. 
These plots reveal the overall performance of the heuristics across the benchmark;
a lower curve means that the heuristic performs better. 

The curves corresponding to the prior work lie above the opt-tree algorithm, i.e., it outperforms 
all the prior algorithms on every tensor in the benchmark.
The performance gain is dependent on the meta-data. 
and varies from 1.5x to 7x. The tensors that achieved the minimum and the maximum gains are: 
Min - $400 \times  400 \times 20 \times 20 \times 20$ compressed to $320 \times 40  \times 10 \times 10 \times 10$;
Max - $400 \times 100 \times 100 \times 50 \times 20$ compressed to $80 \times 80 \times 10 \times 40 \times 10$.
The median improvement is $3.4$x for $5$D and $4.0$x for $6$D tensors.
A detailed study is required to characterize the gain in terms of meta-data.

%We can see that the performance gain on $6$D tensors is better than that of $5$D tensors,
%because there the opt-tree algorithm has more opportunities for rearrangement and reuse of the TTM operations.

We also studied the performance of the algorithms on the real tensors. Figure \ref{fig:real_improvements} shows the actual execution time for one HOOI invocation.
For each tensor, we show 4 bars, corresponding to three prior algorithms and the opt-tree algorithm with dynamic grids. 
For all the tensors, we see that balanced tree outperforms the chain algorithms, because it reuses TTM operations.
The opt-tree algorithm offers improvements as high as 4.6x over (chain, $h$), 5.8x over (chain, $K$) and 4.1x over (balanced). 
For these tensors, the superior performance of the opt-tree algorithm is mainly because of drastic reduction in communication time and partial reduction in computation time. 
Remarkably, the opt-tree algorithm becomes near communication-free under all the three tensors.

%\subsection{Benchmark Performance Analysis}
%\label{sec:expt_analysis}
%In this section, we provide a detailed analysis of the performance of our algorithm on the benchmark tensors. 
%We first study the benefits offered by optimal TTM-tree construction, followed by an analysis of the dynamic gridding scheme.

\subsubsection*{Computation Optimization}
\label{sec:expt_computation_opt}
Here, we study the performance gains from optimal computation tree construction by 
comparing heuristics and the opt-tree algorithm on computation time and load for the TTM-component. 
We normalized the quantities with respect to the opt-tree algorithm.
The time and load for each algorithm-tensor pair was normalized w.r.t the time and load of the opt-tree algorithm. 
The comparison of the time for $5$D and $6$D tensors are reported in Figure \ref{fig:ttm_comp_time_5D} and \ref{fig:ttm_comp_time_6D}.
The opt-tree algorithm offers 1.5-1.7x median improvement compared to prior algorithms for $5$D tensors and 1.4-2.0x median improvement for $6$D tensors. 
The maximum gain is as high as 2.8x and 3.7x for $5$D and $6$D. 

Figure \ref{fig:ttm_comp_load_5D}  and \ref{fig:ttm_comp_load_6D} show the normalized computational load for $5$D and $6$D. 
We see that the opt-tree algorithm offers up to $2.8$x ($5$D) and $3.6$x ($6$D) reduction in load over the best prior algorithm, 
corroborating the improvements seen in time. 
The improvements are higher for $6$D, compared to $5$D,
because opt-tree has more opportunities for careful placement and reuse of the TTMs.

\subsubsection*{Communication Optimization}
\label{sec:expt_communication_opt}
In this experiment, we study the benefits of dynamic gridding.
To do so, we compare the opt-tree algorithm with the static and the dynamic gridding schemes 
under the metrics of communication time and volume.
For the latter, we include the time incurred in TTM multiplication, as well as regridding.
The quantities are normalized with respect to the dynamic gridding scheme.
The results are shown in Figure \ref{fig:comm_time} and \ref{fig:comm_load}. 
In Figure \ref{fig:comm_load}, we can see that dynamic gridding offers up to $6$x factor improvement in communication volume over static gridding,
whereas in Figure \ref{fig:comm_time}, we can see improvements up to $17$x factor (median $9.4$x) in communication time.
The reason for higher improvements on communication time is that regridding (based on all-to-all collective)
turns out to be faster than TTM multiplication (based on reduce-scatter over group communicators) for the same communication volume.
Remarkably, the dynamic grid scheme outperforms static grid scheme on almost all the tensors in the benchmark, with a gain of at least $3$-factor on $90$\% of the tensors.
The gain in communication time is a result of improvement in communication volume, a machine independent statistic.
Thus, we expect similar gains on other distributed memory systems as well.

% Prakash: add a line about comm being killed for 75\% tensors?

% \subsection{Performance on Real Tensors}
% \label{sec:expt_real}
% \begin{figure}[h]
% \centering
% \includegraphics[scale=\pltscale]{expt_figures/RealTensors/RealComparison/bar_chart.pdf} 
% \label{fig:real_improvements}
% \caption{Overall improvements in HOOI iteration time for real tensors. CK:(chain,$K$), CH: (chain, $h$), BN: (balanced, naive), OPT:(opt-tree, dynamic grid)}
% \label{fig:real_improvements}
% \end{figure}

\eat{
\subsubsection{Scaling study}
\label{sec:scaling}

We present a preliminary scaling study of our algorithm on the real tensors on $32$ to $128$ nodes. The execution times are reported in table below. We can see that the TTM component scales well with increasing number of nodes. 
\begin{table}[h!]
\centering
\begin{tabular}{l|lll}
Ranks & 32   & 64  & 128 \\ \hline
HCCI  & 3.9  & 2.1 & 1.4 \\
TJLR  & 10.0 & 5.1 & 3.4 \\
SP    & 5.2  & 2.8 & 1.4
\end{tabular}
%\caption{Execution time of opt-tree algorithm in seconds}
%\label{tab:scaling}
\end{table}
}

\section{Conclusions}
We studied the Tucker decomposition for dense tensors for the distributed memory setting.
We proposed efficient algorithms for computing the optimal trees and dynamic gridding schemes.
Our experimental evaluation on a large benchmark demonstrates that the proposed algorithms
lead to significant reduction in computational load and communication volume,
and offers up to $7$x improvement in performance.
To further improve the performance of HOOI on dense tensors, a distributed SVD solver could be used
instead of the Gram product followed by sequential EVD. Investigating the applicability of the techniques developed
in this paper to the case of sparse tensors is a potential avenue of future work.

\paragraph{Acknowledgements}
We thank Woody Austin, Grey Ballard and Tamara G. Kolda for sharing their insights with us, and the reviewers for helpful comments.

\bibliographystyle{IEEEtran}
\bibliography{main}

\end{document}